\documentclass{aastex}
\usepackage{CJKutf8}
\usepackage{amsmath}
\usepackage{xspace}
\usepackage{graphicx}
\usepackage{epstopdf}
\usepackage{placeins}

\shorttitle{CO overtone IGRINS observation}
\shortauthors{Lee et al.}

\begin{document}
\title{IGRINS spectroscopy of Class I sources: IRAS~03445+3242 and IRAS~04239+2436}


\author{Seokho Lee, Jeong-Eun Lee, Sunkyung Park}
\affil{School of Space Research, Kyung Hee University, Yongin-shi, Kyungki-do 449-701, Korea}
\email{jeongeun.lee@khu.ac.kr}
\author{Jae-Joon Lee}
\affil{Korea Astronomy and Space Science Institute 776, Daedeokdae-ro, Yuseong-gu, Daejeon, 305-348,  Korea}
\author{Benjamin Kidder, Gregory N. Mace, \and Daniel T. Jaffe}
\affil{Department of Astronomy, University of Texas, R.L. Moore Hall, Austin, TX 78712, USA}

\begin{abstract}We have detected molecular and atomic line emission from the hot and warm disks of two Class I sources, IRAS~03445+3242 and IRAS~04239+2436 using the high resolution Immersion GRating INfrared Spectrograph (IGRINS). CO overtone band transitions and near-IR lines of Na I and Ca I, all in emission, trace the hot inner disk while CO rovibrational absorption spectra of the first overtone transition trace the warm gas within the inner few AU of the disk. The emission-line profiles for both sources show evidence for Keplerian disks.  A thin Keplerian disk with power-law temperature and column density profiles with a projected rotational velocity of $\sim$60--75~km~s$^{-1}$ and a gas temperature of $\sim$3500~K at the innermost annulus can reproduce the CO overtone band emission.
Na~I and Ca~I emission lines also arise from this disk, but they show complicated line features possibly affected by photospheric absorption lines. Multi-epoch observations show asymmetric variations of the line profiles on one-year (CO overtone bandhead and atomic lines for IRAS~03445+3242) or on one-day (atomic lines for IRAS~04239+2436) time scales, implying non-axisymmetric features in disks.
The narrow CO rovibrational absorption spectra ($v$=0$\rightarrow$2) indicate that both  warm ($>$ 150~K) and cold ($\sim$20--30~K) CO gas are present along the line of sight to the inner disk. This study demonstrates the power of  IGRINS as a tool for  studies of the sub-AU scale hot and AU-scale warm protoplanetary disks with its simultaneous coverage of the full H and K bands with  high spectral resolution ($R$= 45,000) allowing many aspects of the sources to be investigated at once. 
\end{abstract}
%
%
%
%
\section{INTRODUCTION }\label{sec:intro}
 The physical and chemical structure of the inner gaseous disks around young stellar objects is very important to an understanding of the environment in which  planets form, as well as to an understanding of accretion processes in star formation. The timescale for gas dissipation in the inner disk will determine which formation mechanisms for giant planets, i.e., gravitational instability \citep{Boss1997, Mayer2002} or core accretion \citep{Bodenheimer2002} are plausible. Terrestrial planet formation is also very sensitive to the residual gas mass in the terrestrial zone at ages of a few million years, since the final mass and eccentricities that are important for habitability strongly depend on the gas column density in the region \citep{Kominami2002}. 
 
 High spectral resolution spectroscopy of the many atomic and molecular lines in the near infrared (IR) provides useful information about the  physical and chemical conditions in the inner gaseous disks surrounding young stellar objects. 
 CO overtone ($\Delta v= 2$) emission arises in hot (1000 K--5000 K) and dense ($\ge 10^{10} {\rm cm^{-3}}$) regions, and is one of the most powerful tools for exploring the inner hot disk. The CO overtone bandhead emission has been used to probe  Keplerian disks \citep[e.g.,][]{Carr1993,Najita1996}.  Emission in the near-IR Na I and Ca I lines, also likely arising  in the disk, is correlated with CO overtone bandhead emission \citep{Davis2011,Doppmann2008}.
 
 The CO overtone band emission is more frequently detected in Class I sources \citep[15--25\%,][]{Connelley2010,Doppmann2005} than Class II, Classical T Tauri stars \citep[a few \%,][ hereafter CTTSs]{Carr1989,Greene1996}.
In Herbig Ae/Be (HAeBe) stars, higher mass counterparts of CTTSs, the CO overtone band emission was detected only  $\sim$7\% \citep{Ilee2014,van der Plas2015}.
 However, in Class I protostars, the modeling of the CO overtone band emission has been done only for the most luminous protostars; WL~16 ($L_{\rm bol}\sim$22 $L_{\odot}$) is the lowest luminosity Class I source \citep{Najita1996} where detailed modeling of the high resolution CO overtone bandhead emission has been performed.
The modeling for CTTSs and HAeBe stars indicates the CO overtone emission arises in the inner $\sim$0.15~AU Keplerian disks \citep[e.g.,][]{Berthoud2007,Najita1996,Najita2009}.
Interferometric observations toward CTTSs and HAeBe stars show that the CO overtone band emission has a spatial distribution smaller than or similar to the distribution of the continuum emission \citep{Eisner2014,Tatulli2008}. 
  
 The Immersion Grating INfrared Spectrograph \citep[IGRINS,][]{Yuk2010,Park2014} has a resolving power  $R$= $\lambda/\Delta\lambda= 40,000-50,000$  and covers the full H and K bands in the near-IR. 
 Therefore the instrument covers multiple species and multiple transitions in a single exposure and  provides a broad picture of the kinematics and physical conditions of a target. 
 This capability of IGRINS enables us to carry out a detailed study of the CO overtone band spectra from protoplanetary disks.  The spectra in the bandhead and ``bandtail" ($\sim$2.3~$\mu$m) of CO $v$=2--0 trace the innermost and outermost parts of the hot CO gas disk, respectively; the projected
velocity at the innermost part can be derived from the shape of bandhead while the projected velocity at the outermost part can be derived from the velocity difference between the double peaks of those isolated rovibrational lines in ``bandtail"  \citep[e.g.,][]{Najita1996,Smak1981}. In addition, we can test whether the gas emitting the CO overtone bands is vibrationally thermalized by comparing the overtone band emission over the range from $v$=2--0 to $v$=5--3.

Furthermore, the broad wavelength coverage of IGRINS combined with such a high spectral resolution provides an opportunity to detect the narrow absorption features of CO $v=0 \rightarrow 2$, which arise either from the disk or a protostellar envelope. There are only a few studies on the narrow CO overtone absorption features toward Class I sources and CTTSs with NIRSPEC ($R\sim25,000$) on Keck and CRIRES ($R\sim95,000$) on VLT \citep{Horne2012,Smith2009,Smith2015}.

 We observed IRAS~03445+3242 and IRAS~04239+2436  as part of an IGRINS survey of disks
 \footnote{http://kgmtscience.kasi.re.kr/kgmt\_igrins/legacy/selected.html} 
 (PI. J.-E. Lee) in sources with a wide range of ages, luminosities, and masses (from Class I to weak T Tauri stars, and from low-mass CTTSs to HAeBe stars and massive young stellar objects). 
 While prior studies of CO overtone band emission have been performed at medium resolution \citep{Connelley2010} or have focused on higher luminosity sources \citep{Najita1996}, this is the first high resolution near-IR spectroscopic observation toward low luminosity ($L_{\rm bol} < 5~L_\odot$) Class I sources. The two targets are nearby Class I sources associated with reflection nebulae. According to \citet{Connelley2010} and \citet{Davis2003}, these two sources have high veiling ($r_{\rm K} > 5$), so that photospheric absorption lines from the protostars themselves have very small equivalent widths. 
 
In this paper, we analyze lines detected in high resolution IGRINS spectra, especially the CO overtone bands, Na I 2.2 $\mu$m doublets, and Ca I 1.978 and 1.987 $\mu$m lines, toward IRAS~03445+3242 and IRAS~04239+2436 using a simple Keplerian disk model. We also analyze the CO $v=0\rightarrow2$ overtone absorption features toward both sources. The combination of these high-resolution CO rovibrational emission and absorption lines provides us a unique opportunity to study protoplanetary disks in the temperature range of a few hundreds K to above the dust sublimation temperature.

\section{OBSERVATIONS AND DATA REDUCTION}\label{sec:obs}
We observed IRAS~03445+3242 and IRAS~04239+2436  with IGRINS on the 2.7~m Harlan J. Smith Telescope of the McDonald observatory at multiple epochs. Table~\ref{tb:obslog} lists a log of our observations.
The data were taken in a series of ABBA nodding observations with an offset of 7$''$ along the $15''\times1''$ slit. 
 The data are reduced using the IGRINS pipeline version 2.1\footnote{https://github.com/igrins/plp; https://zenodo.org/record/18579\#.VX7zPeuTrlc}. The raw data are flat-fielded and flagged for bad pixels. The one-dimensional spectra are extracted from pair-subtracted images (A-B) using a modified version of the optimal extraction algorithms of \citet{Horne1986}.  The wavelength solutions were derived using OH lines from spectra of blank sky and telluric absorption lines from the source spectra. Nearby A0 stars are observed as telluric standards. A telluric correction is applied after scaling and shifting the A0 spectra to compensate for  differences in airmass and changes in the wavelength solution. 
   We calibrate the absolute flux by scaling the spectroscopic continuum to the 2MASS magnitude of each source.  The continuum emission is fit with a polynomial over the whole K-band and subtracted from the spectra.

\section{RESULTS AND ANALYSIS}\label{sec:model}
\begin{figure*}
\includegraphics[width=1 \textwidth]{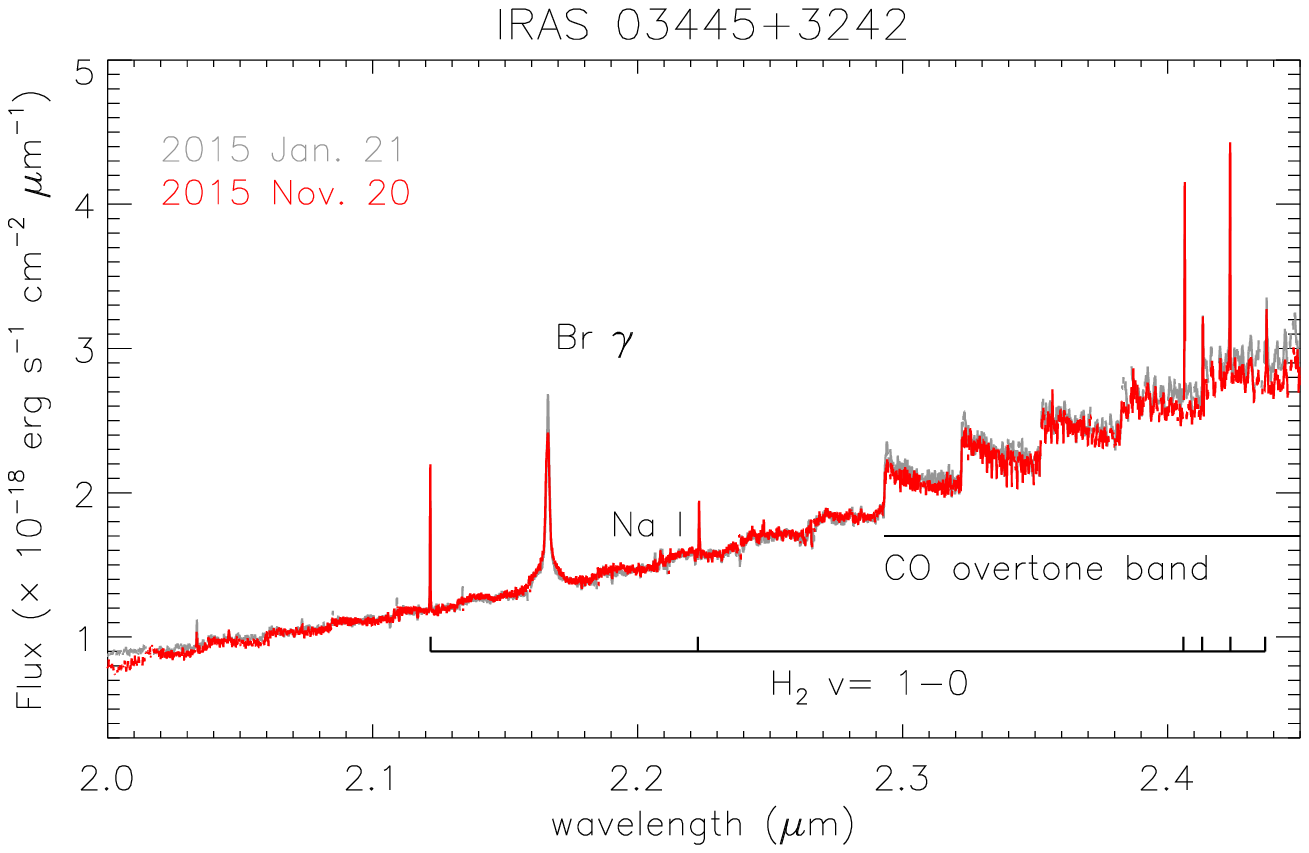}
\includegraphics[width=1 \textwidth]{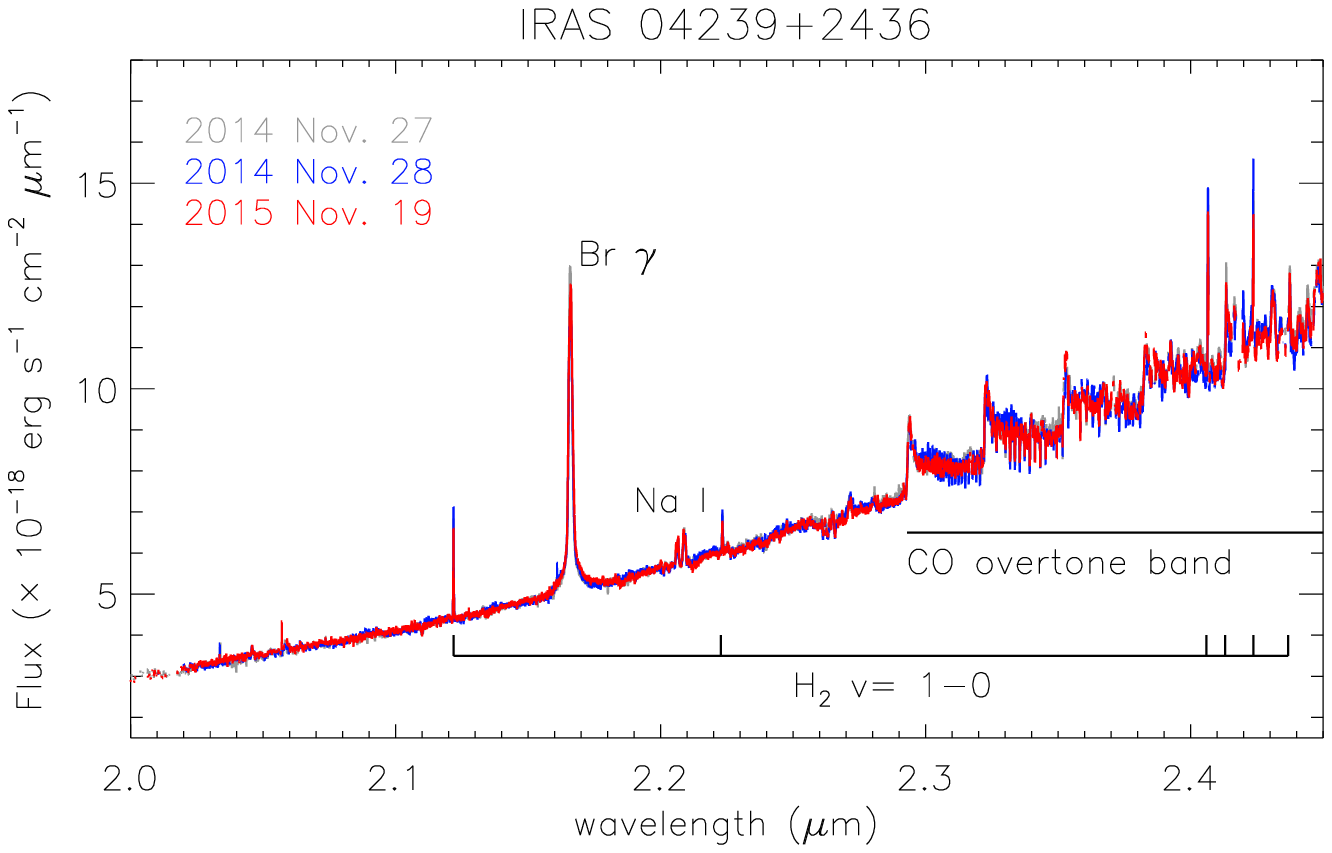}
\caption{K-band spectra of IRAS~03445+3242 (top) and IRAS~04239+2436 (bottom). Different colors indicate different epochs for observations.}
\label{fig:kband}
\end{figure*}

 IRAS~03445+3242 and IRAS~04239+2436 show many emission lines in the H and K bands, most of which were already seen in previous \textit{low-}resolution observations \citep[e.g.][]{Connelley2010}.
As shown in Figure~\ref{fig:kband}, Br $\gamma$, molecular hydrogen rovibrational transitions, CO overtone rovibrational lines, and Na I lines are all detected in emission. Many hydrogen recombination lines and [Fe II] lines in the H band and Ca I lines in the short wavelength part of K band are also detected. 

 In the multi-epochs, although each emission line profile varies, the slope of continuum shows no drastic change with time, and no emission lines totally disappear or appear.
The red lines in Figure~\ref{fig:kband} show the spectra observed almost one year after the first observations (grey lines). The blue line in the bottom panel of Figure~\ref{fig:kband} presents the spectrum of IRAS~04239+2436 obtained a day after the first observation.
 IRAS~03445+3242 shows a photometric decline from 1988 to 1993 without changes in the J-H and H-K colors \citep{Moore1992,Moore1994} while IRAS~04239+2436 shows almost constant photometric magnitude with a small scatter \citep[$\Delta$~K$\le 0.45$~mag,][]{Connelley2014}.
 We note that we did not consider photometric variability in fixing the flux level and assumed that the photometric fluxes are the same as the flux equivalent of the 2MASS magnitude. In addition, line flux densities are measured by subtracting the same continuum level in each epoch. Therefore, line flux densities are relative values to the continuum  when we compare them in multi-epochs.
 In this paper, we focus on the disk tracers such as the CO overtone, Na I, and Ca I emission lines. 
 
 Figure~\ref{fig:co_obs} shows the broadened CO overtone bandhead emission (left) and isolated emission lines in the ``bandtail" of $v$=~2--0 (middle) as well as the narrow $v$=~0$\rightarrow$~2 absorption spectra (right) toward both sources. 
 For IRAS~03445+3242, the peak flux of the bandhead  decreases by $\sim$25\% over 11 months, and some emission peaks in the middle and right panels also decrease. The spectra in the first (grey) and last (red) observations for IRAS~04239+2436 are similar except for the absorption features while the second observation (blue) shows higher emission peaks around 2.295-2.300~$\mu$m (left panel) and 2.2330-2.335~$\mu$m (right panel) than those at other epochs.
 
\subsection{The inner hot disk}\label{ss:in}
\subsubsection{CO overtone band emission}\label{sss:co}
\begin{figure*}
\includegraphics[width=1 \textwidth]{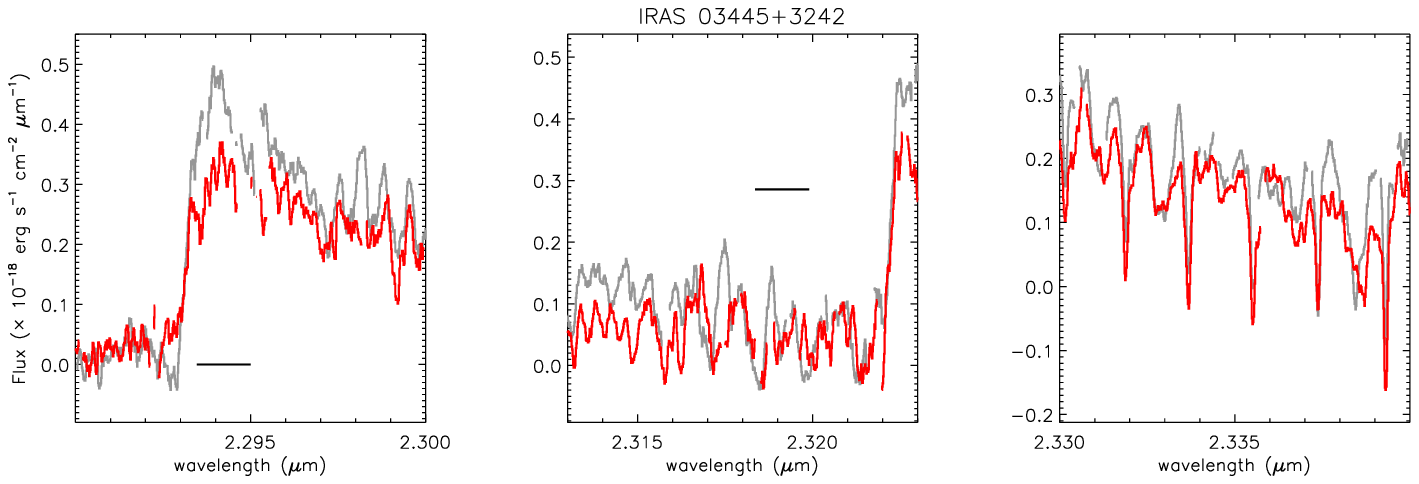}
\includegraphics[width=1 \textwidth]{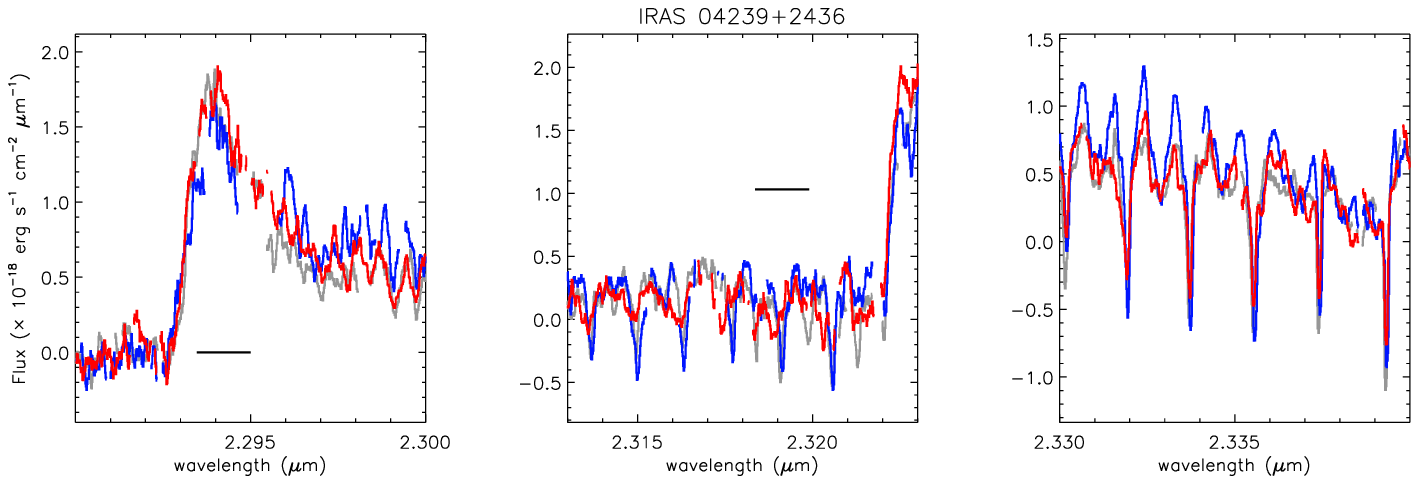}
\caption{Bandhead (left) and ``bandtail" of CO $v=$~2-0 (middle) and absorption features of $v= 0\rightarrow 2$ over the CO $v=$~3-1 overtone band (right) for IRAS~03445+3242 (top) and IRAS~04239+2436 (bottom). The colors are same as Figure~\ref{fig:kband}. The horizontal black lines in the left and middle panels indicate a length scale of 200~km~s$^{-1}$. The observed spectra are boxcar smoothed by 15~km~s$^{-1}$ for clear spectral features.}
\label{fig:co_obs}
\end{figure*}

\begin{figure*}
\includegraphics[width=0.7 \textwidth]{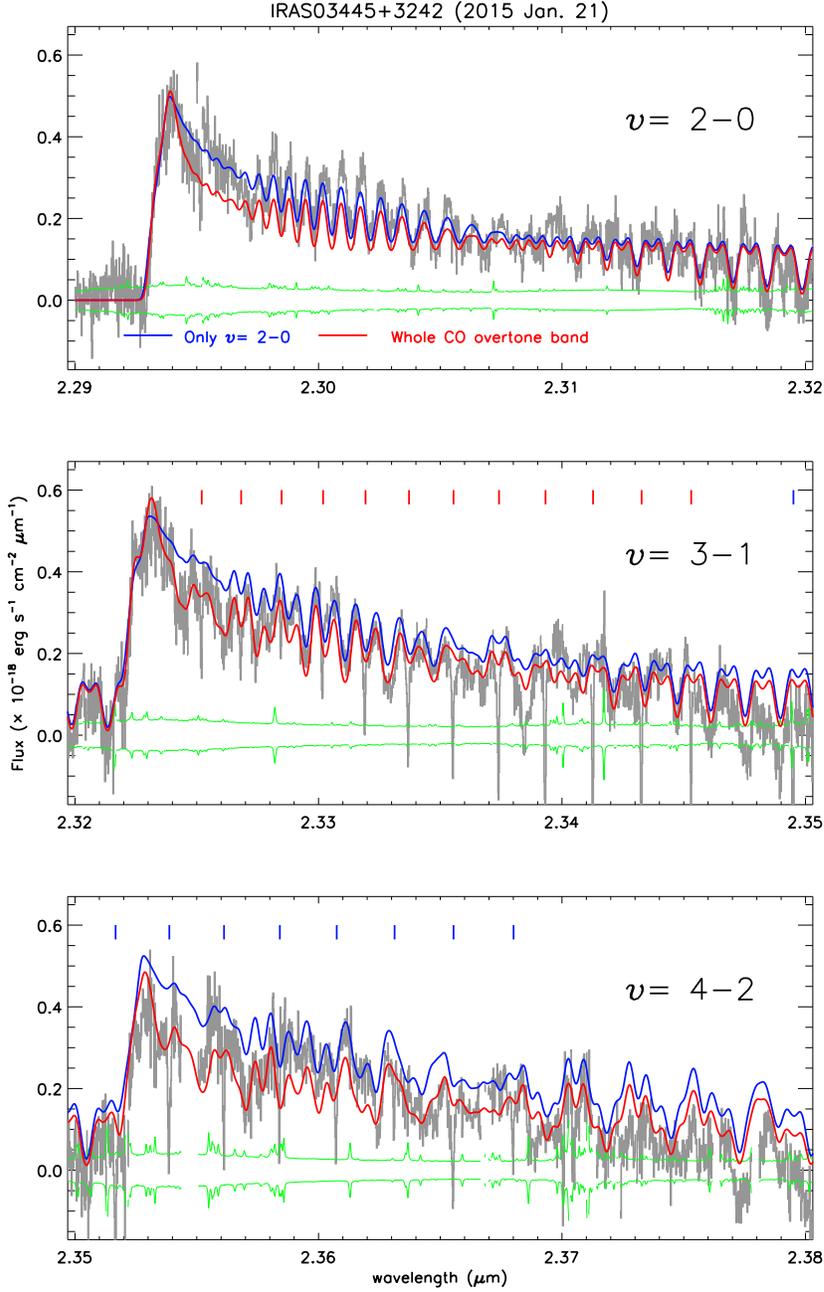}
\caption{The CO overtone band emission spectra of IRAS~03445+3242. The observed spectra on 21 Jan. 2015 are plotted in grey, and the synthesized spectra of the best-fit models for the whole CO overtone band (Table~\ref{tb:bestfit}) and for only $v$=2--0 (Table~\ref{tb:bestfit1}) are plotted in red and blue lines, respectively.  The green lines indicate the $\pm$1 $\sigma$ observation errors.
The sharp features in the green lines are caused by telluric contamination. The red and blue tick marks  indicate the position of narrow absorption R- and P-branch rovibrational transitions of $v$=~0$\rightarrow$2.   }
\label{fig:co_om-i03-1}
\end{figure*}

  In both IRAS~03445+3242 and IRAS~04239+2436, the CO overtone $v$= 2--0 bandhead emission (left panels in Figure~\ref{fig:co_obs}) shows ``blue wing", ``shoulder", and ``red peak" features, as seen in WL 16 \citep{Carr1993,Najita1996}.  In addition, isolated rovibrational transitions in the ``bandtail" of the CO $v$=2--0 band (middle panels of Figure~\ref{fig:co_obs} and the top panel of Figure~\ref{fig:co_om-i03-1}) show double peaks .  These features have been interpreted in sources with CO overtone emission as evidence for a Keplerian disk  \citep[e.g.,][]{Carr1993,Najita1996}.
 
The characteristic shapes of the CO features lead us to investigate whether the CO bands can  be produced by emission from a thin Keplerian disk. We assume  power-law distributions of the gas temperature $T(r)$ and the CO column density $N(r)$:
\begin{eqnarray}
\label{eq:disk_TN}
T(r) & =  &T_{\rm in} \cdot \left(\frac{r}{r_{\rm in}}\right)^{-p} \quad \rm{K} \\
N(r) & =  &N_{\rm in} \cdot \left(\frac{r}{r_{\rm in}}\right)^{-q} \quad \rm{cm^{-2}},
\end{eqnarray}
where $T_{\rm in}$ and $N_{\rm in}$ are the gas temperature and the CO vertical column density at the innermost annulus $r_{\rm in}$. $p$ and $q$ are fixed to 0.5 and 1.5, respectively, following \citet{Eisner2014}. CO is dissociated above 5000 K and CO overtone bandhead emission can be produced only in high temperature gas ($>$ 1000 K). We explored values of $T_{\rm in}$  and $T_{\rm out}$ (the gas temperature at the outermost annulus $r_{\rm out}$) from 1000 K up to 5000 K ($T_{\rm in}\ge T_{\rm out}$).

At each annulus, we calculate the CO band emission using $I_\lambda(r)= B_\lambda(T(r))\left(1 - \exp(-\tau_\lambda)\right)$. The optical depth $\tau_\lambda$ is calculated using a given Gaussian profile with a line width ($\Delta V_{\rm FWHM}$), $T(r)$, and $N(r)$  assuming local thermal equilibrium (LTE). The CO molecular data are adopted from the HITRAN database \citep{Rothman2013}. The spectrum is then convolved with the rotational broadening function \citep[e.g.,][]{Calvet1993}:  
\begin{equation}
\Psi (\Delta V) = \left[ 1 - \left(\frac{\Delta V}{V_{\rm max}}\right)^2 \right]^{-1/2} 
\quad ( -V_{\rm max} \le \Delta V \le V_{\rm max})
\end{equation}
where $\Delta V$ is the velocity shift from the line center. The maximum projected rotational velocity is $V_{\rm max} = V(r) \sin i$, where $V(r)$ is the Keplerian velocity at the annulus located at $r$ and $i$ is the inclination. 
The emission from each annulus is weighted by its area. Next, the emission from the annuli is summed from $r_{\rm in}$ to $r_{\rm out}$ and  convolved with the instrument resolution ($\sim$7.5~km~s$^{-1}$). We then correct for the radial velocity of each source. 
We adopted the source velocities, $V_{\rm LSR}$, of 10.27 $\pm$ 0.06 km~s$^{-1}$ derived from  N$_2$H$^+$~1--0 observations for IRAS~03445+3242 \citep{Emprechtinger2009} and 6.54 $\pm$~0.02~km~s$^{-1}$ from C$^{17}$O~1--0 observations for IRAS~04239+2436 \citep{Fuller2002}.

 Lastly, we corrected the extinction using a  simple power law  $A_\lambda=~A_{\rm V} \left(0.55~\mu m/\lambda\right)^{1.6}$ \citep{Rieke1985} and the visual extinction, $A_{\rm V}$, for each source listed in Table~\ref{tb:target}. Observations toward these sources with the Infrared Spectrograph aboard the Spitzer Space Telescope show 9.7 $\mu$m silicate absorption features \citep{Boogert2008}. The silicate optical depth, $\tau_{9.7}$, toward IRAS~03445+3242 and IRAS~04239+2436 is 1.23 and 1.05, respectively, with a 10\% uncertainty. The visual extinction is derived from $A_{\rm V}/\tau_{9.7} = 18.5$ \citep{Roche1984}. Note that our values are lower limits because $\tau_{9.7}$ underestimates $A_{\rm V}$ in dense clouds \citep{Chiar2007}. 

 The best-fit source models are found by $\chi^2$ minimization  varying the model parameters of  $V_{\rm in}$ (=$V(r_{\rm in})\sin i$), $\Delta V_{\rm FWHM}$, $T_{\rm in}$, $T_{\rm out}$, and log($N_{\rm in}$) in steps of 5~km s$^{-1}$, 5~km s$^{-1}$, 50~K, 500~K, and 0.3, respectively. 
 The solid angle of the source, $\Omega$ is defined as 
\begin{equation}
\Omega = \pi \left(\frac{r_{\rm in}}{d}\right)^2 \cos i \left(\frac{T_{\rm in}}{T_{\rm out}} -1\right),
\end{equation} 
where d is the distance to the source.  The solid angle  is calculated to minimize $\chi^2$ ($\partial \chi^2/\partial \Omega  = 0$) in order to match the observed flux.

\begin{figure*}
\includegraphics[width=0.7 \textwidth]{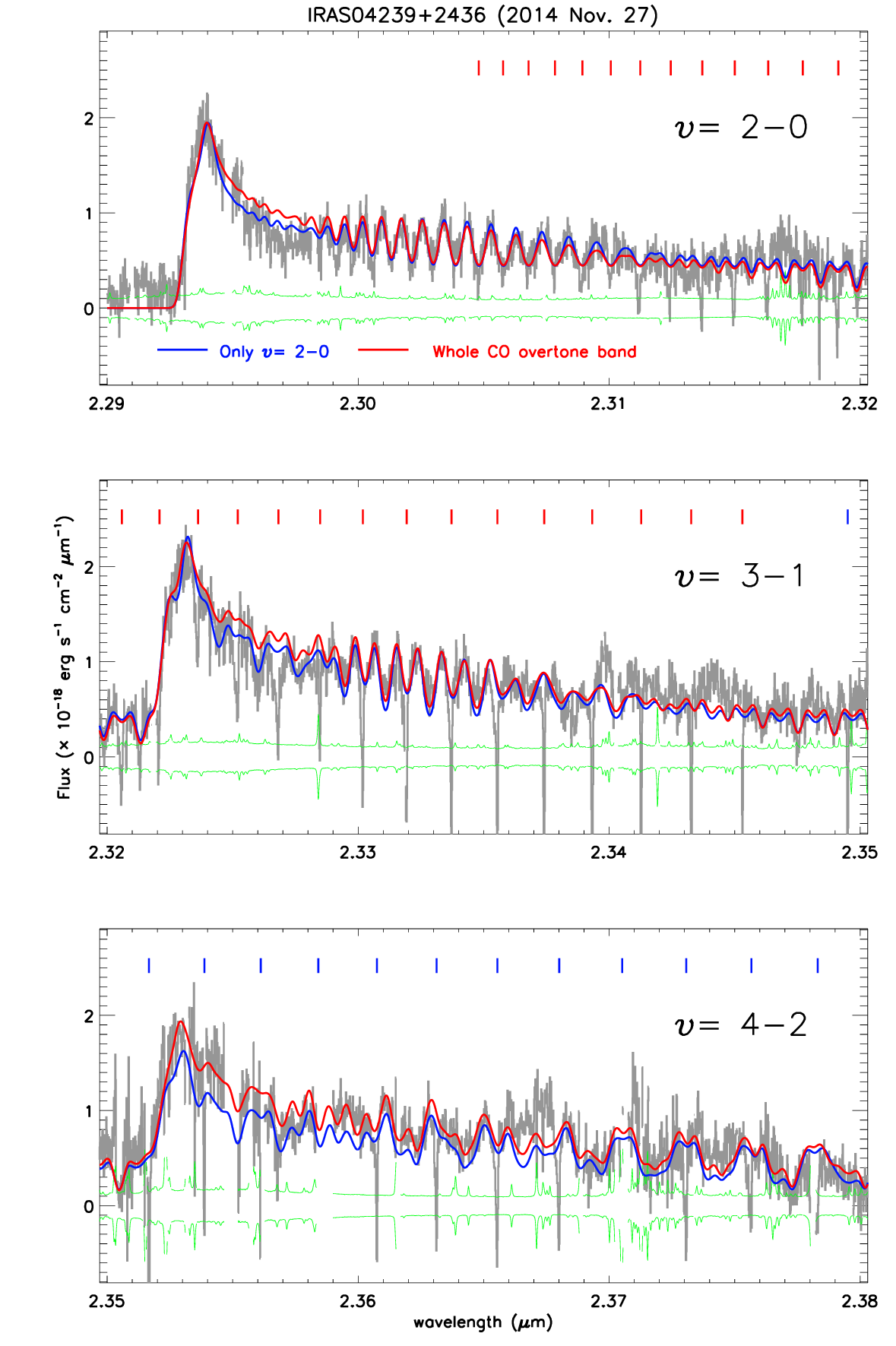}
\caption{Same as Figure~\ref{fig:co_om-i03-1} except for the CO overtone band emission spectra of IRAS~04239+2436 observed on 27 Nov. 2014.}
\label{fig:co_om-i04-1}
\end{figure*}

\begin{figure*}
\includegraphics[width=0.7 \textwidth]{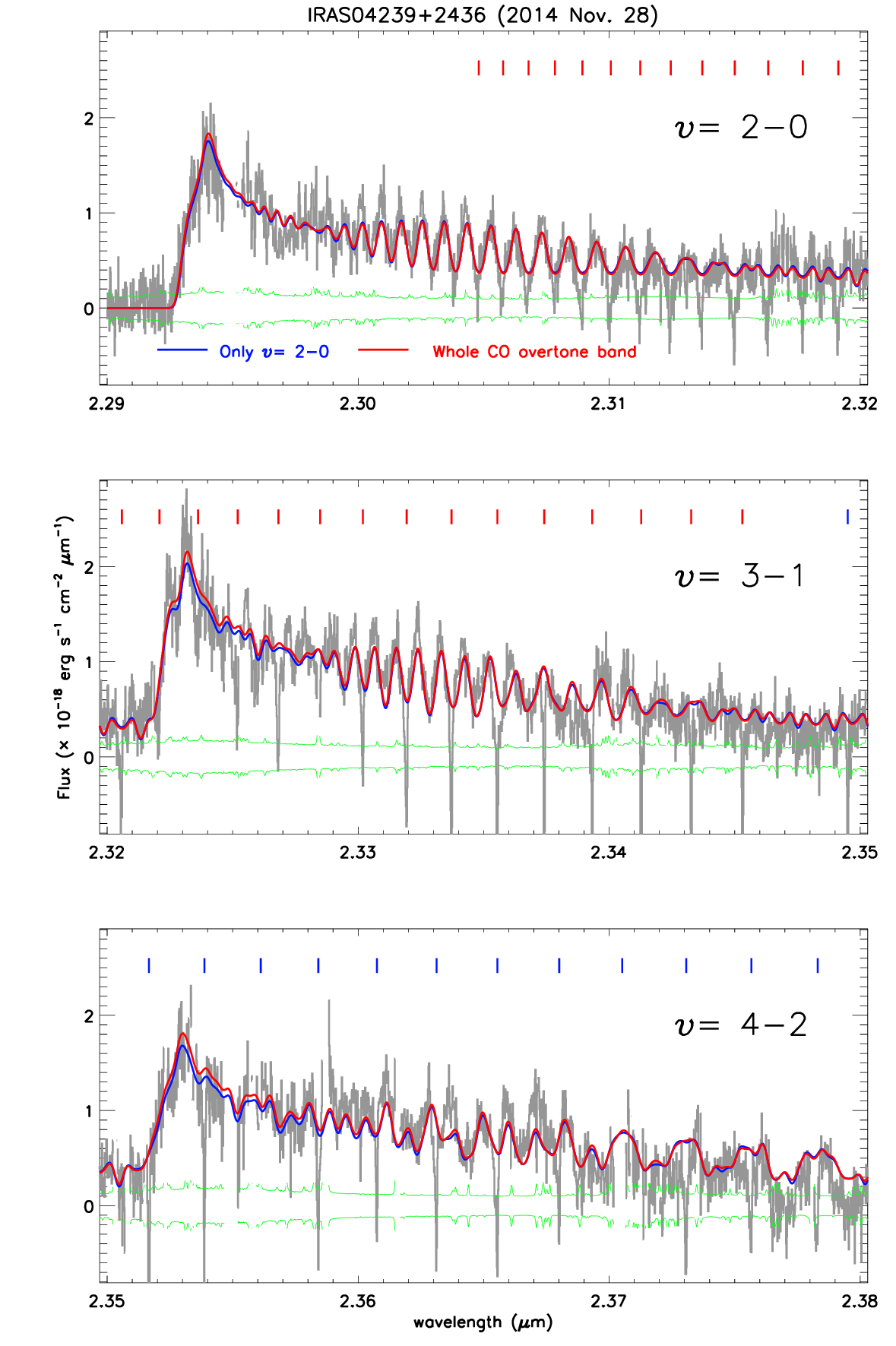}
\caption{Same as Figure~\ref{fig:co_om-i03-1} except for the CO overtone band emission spectra of IRAS~04239+2436 observed on 28 Nov. 2014. }
\label{fig:co_om-i04-2}
\end{figure*}

\begin{figure*}
\includegraphics[width=0.7 \textwidth]{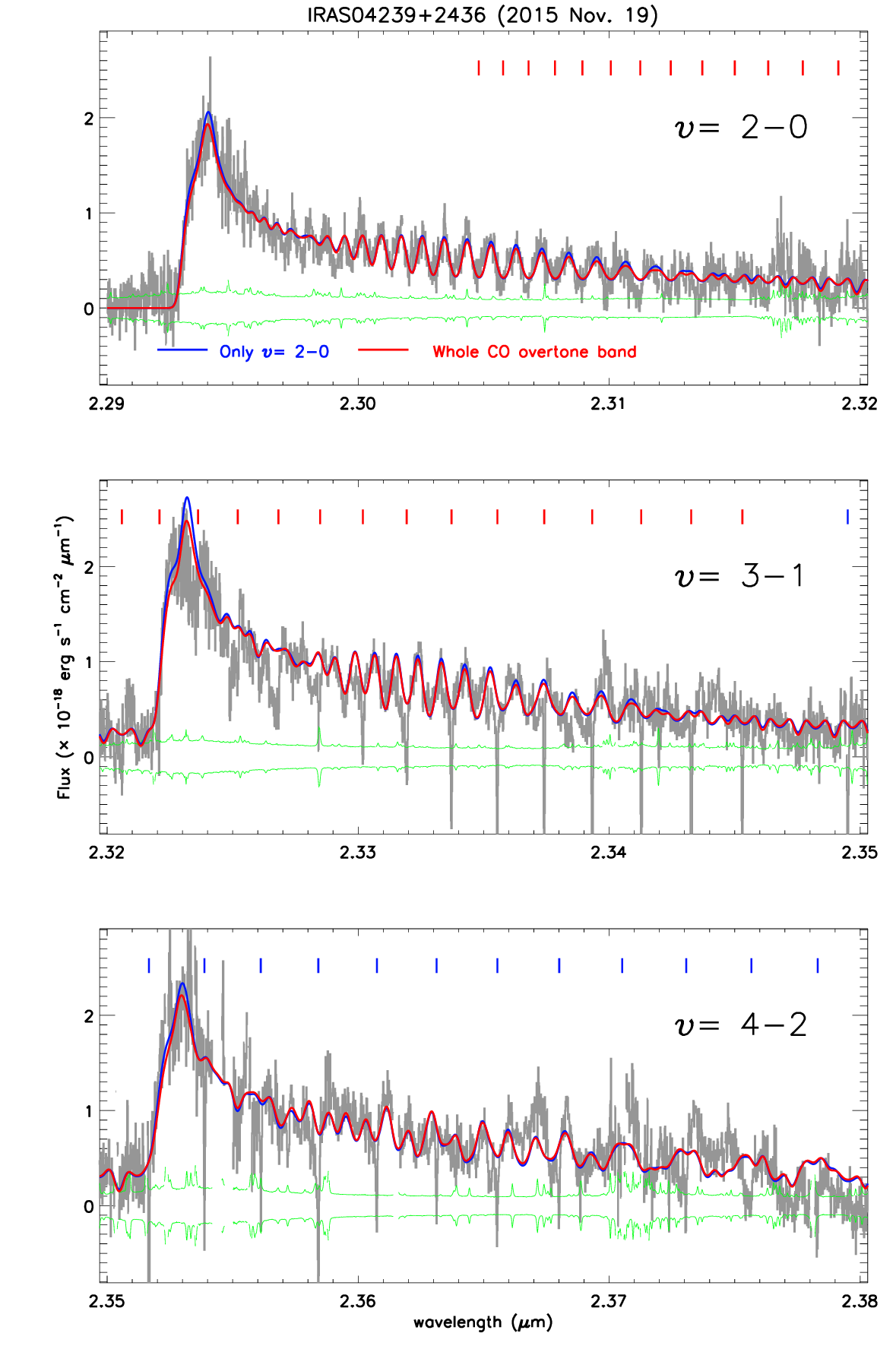}
\caption{Same as Figure~\ref{fig:co_om-i03-1} except for the CO overtone band emission spectra of IRAS~04239+2436 observed on 19 Nov. 2015.  }
\label{fig:co_om-i04-3}
\end{figure*}

The best-fit models (Table~\ref{tb:bestfit}) for all of the CO overtone transitions covered by  IGRINS show a narrow annulus with $T_{\rm in}$ of 3,000--3,750~K and $T_{\rm out}$ of 2,500--3,000~K for both sources.
The projected Keplerian velocity $V_{\rm in}$ is $\sim$60~km~s$^{-1}$ and $\sim$75~km~s$^{-1}$ for IRAS~03445+3242 and IRAS~04239+2436, respectively (see the red lines in  Figures~\ref{fig:co_om-i03-1}--\ref{fig:co_om-i04-3}), and the line-width $\Delta V_{\rm FWHM}$ is $\sim$50~km~s$^{-1}$ for the both sources. 

 When we adopt the inclination from Table~\ref{tb:target}, we derive an inner radius $r_{\rm in}$ of 0.024--0.039 (0.021--0.031)~AU and a protostellar mass $M_{\rm star}$ of 0.14--0.17 (0.14--0.20)~$M_\odot$ for IRAS~03445+3242 (IRAS~04239+2436) from the solid angle of the source and $V_{\rm in}$ \citep[e.g.,][]{Berthoud2007}. The derived disk inner radius is 3--5 times larger than the protostellar radius ($\sim$0.008~AU) for the derived protostellar masses when the protostars are at the birthline with a mass accretion rate of $\dot{M}=2\times10^{-5}~{\rm M}_\odot {\rm yr}^{-1}$ \citep{Hartmann1997}. In CTTSs and HAeBe stars, interferometric observations show that the CO overtone emission arises inside of $\sim$0.15~AU \citep{Eisner2014}. In our model, most bandhead emissions arises within $\sim$0.05~AU for both sources.

A dust sublimation radius is calculated as \citep{Monnier2002}
\begin{equation}
R_{\rm S} = 7.5 \sqrt{Q_{\rm R}} \sqrt{L_\star/L_\odot} \left(\frac{T_{\rm sub}}{1500~{\rm K}}\right)^{-2} R_\odot,
\end{equation}
where $Q_{\rm R}$ is the ratio of the dust absorption efficiencies  at the color temperatures of the incident and re-emitting radiation fields and $T_{\rm sub}$ is the dust sublimation temperature. When we adopt a stellar luminosity $L_\star$ as the bolometric luminosity in Table~\ref{tb:target}, $Q_{\rm R}$=~4, and $T_{\rm sub}$=~1500~K \citep{Dullemond2001}, the dust sublimation radius is 0.14 and 0.07~AU for IRAS~03445+3242 and IRAS~04239+2436, respectively. Therefore, most CO overtone band emission is radiated from the dust-free inner disk in both sources.

\subsubsection{Na I and Ca I emission} \label{sss:atom}
The behavior of the 2.2 $\mu$m Na I lines and the Ca I lines at 1.978 and 1.987~$\mu$m, all of which are prominent photospheric absorption lines in late-type stellar atmospheres, correlates with the behavior of the CO overtone transitions, which are also prominent absorption features in cool stellar atmospheres  \citep{Antoniucci2008,Davis2011}; these lines are detected as emission (absorption) when the CO overtone bandhead  is detected in emission (absorption). As expected from the CO overtone band emission, these lines are detected in emission toward our two targets. 
 However, at the high spectral resolution of IGRINS, the combined effect of stellar photospheric absorption lines and emission lines are clearly seen.

\begin{figure*}
\includegraphics[width=0.8 \textwidth]{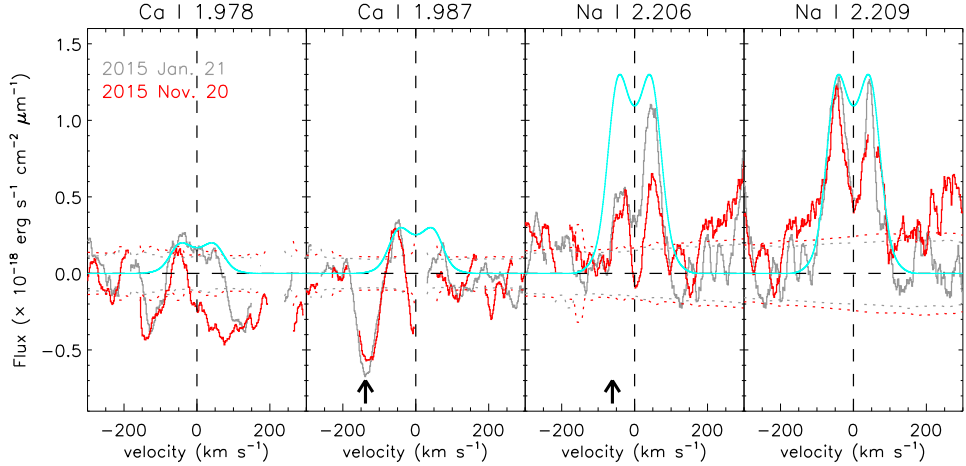}
\caption{IGRINS spectra of Ca I lines at 1.978 and 1.987 $\mu$m and Na I lines at 2.206 and 2.209 $\mu$m observed in different epochs (grey and red colors indicate two epochs as presented in the top of the first panel.) and the synthesized spectra for the best-fit Keplerian disk model  (cyan line, see Section~\ref{sss:atom}) for IRAS~03445+3242. The horizontal axis is the velocity relative to the systemic velocity in Table~\ref{tb:target}. The observed spectra are boxcar smoothed by 15~km~s$^{-1}$ to present the spectral features clearly. The vertical arrows in the second and third panels indicate the wavelength positions of Ca~I~1.986~$\mu$m and Sc~I~2.206~$\mu$m, respectively. The dotted lines represent the observation errors ($\pm$3$\sigma$). Note that the wavelength of atomic lines is adopted from \citet{Rayner2009}.}
\label{fig:atom-i03}
\end{figure*}

 For IRAS~03445+3242, the Ca I and Na I lines in Figure ~\ref{fig:atom-i03} show evidence for time variability that alters the relative contribution of emission lines and  photospheric absorption lines. We can see the Ca~I 1.986~$\mu$m absorption near -138~km~s$^{-1}$ of Ca~I 1.987~$\mu$m (the vertical arrow in the second panel) and absorption features in both side of Ca I 1.978~$\mu$m.  Furthermore, the flux of Na~I 2.206~$\mu$m is lower than that of Na I 2.209~$\mu$m  although both components of the Na I doublet should have similar fluxes and line shapes. Especially, Na I  2.206~$\mu$m has a blue shoulder in the epoch of 21 Jan. 2015  (grey lines). The blue shoulder in Na I 2.206~$\mu$m and strong dips in both Na I doublet are possibly due to the Sc~I 2.206 $\mu$m photospheric absorption  line (the vertical arrow in the third panel) and to photospheric absorption in the Na I doublet lines themselves as shown in GV Tau \citep{Doppmann2008}.
 
Interestingly, the fluxes of the red peaks ($\sim$50~km~s$^{-1}$) of both Ca~I and Na~I lines decrease in one year (the red side of the Ca I 1.987 $\mu$m line is affected  by a poor telluric correction). The height of the red peak of the CO $v$=2--1 bandhead also decreases over a year (see Figure~\ref{fig:co_obs}). This decrease might imply that the density or gas temperature in the inner disk with a projected rotation velocity  of $\sim$50~km~s$^{-1}$ drops.

\begin{figure*}
\includegraphics[width=0.8 \textwidth]{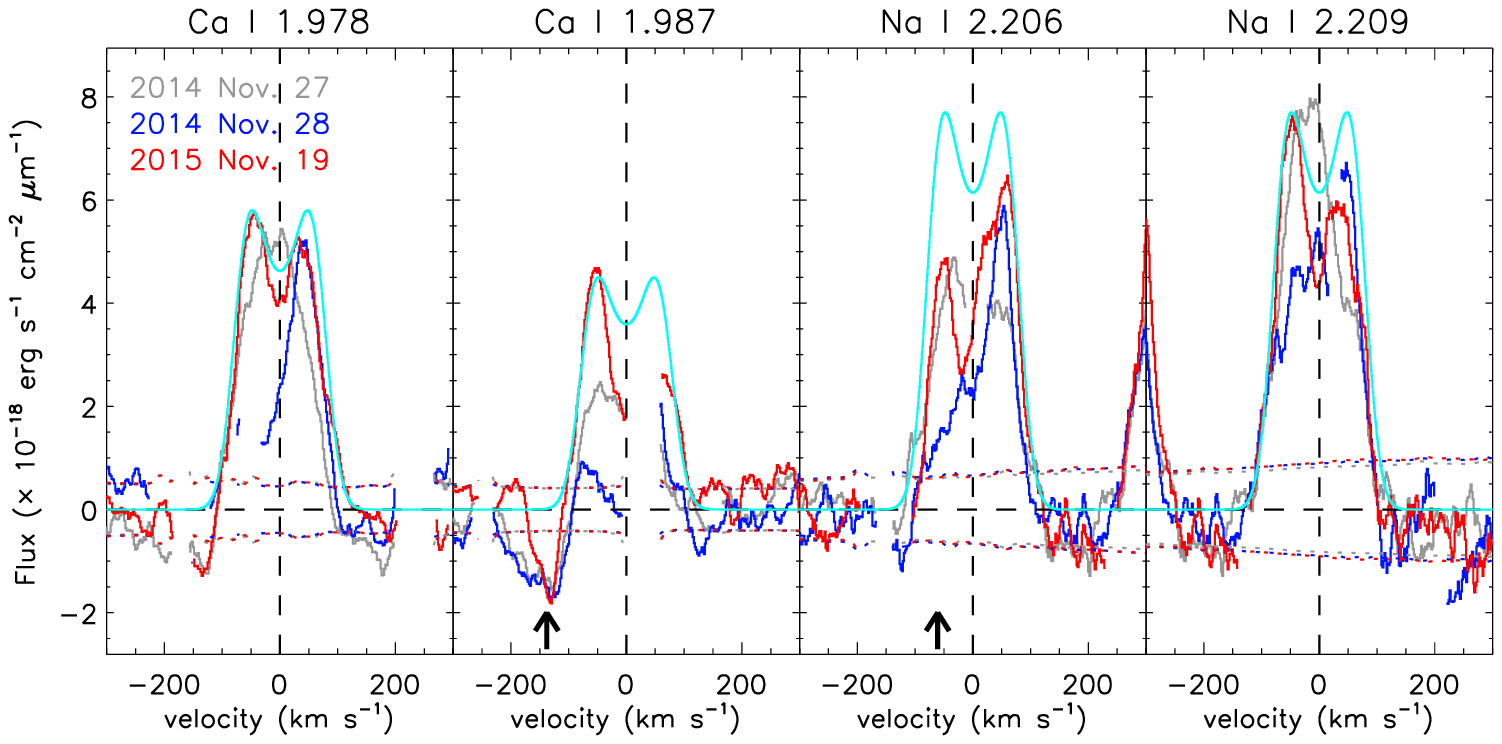}
\caption{Same as Figure~\ref{fig:atom-i03} except for IRAS~04239+2436. }
\label{fig:atom-i04}
\end{figure*}

For IRAS~04239+2436, the spectra of Ca~I and Na~I also vary with time, particularly notable are the strong central dips and the enhancement of the red-peaks that appear over one year (compare the grey and red lines in Figure~\ref{fig:atom-i04}).
 The spectral lines observed in the second epoch (blue lines), which is just one day after the first epoch, show a blue shoulder. The highest velocity components (located at the innermost emitting disk) are almost constant over one year, while the other velocity components vary. 
 Since the two Na I lines should have similar shapes and fluxes, the difference seen in the Na I doublet might be a result of stellar atmosphere. In addition, this source is a binary system \citep{Connelley2008, Reipurth2000}, and thus, two different photospheric spectra (with different line ratios and radial velocities) might affect the atomic lines.

The photospheric absorption lines from the binary system cannot explain the rapid decrease of the flux of the blueshifted velocity component. 
This rapid variation can be caused by a rotation of the protostar \citep[e.g.,][]{Gullbring1996,Herbst1994}.
One possibility is that the blueshifted (by $\sim$50~km~s$^{-1}$) gas in the inner disk surface, which is passively heated, cools down rapidly due to a dark spot on the protostar. If this dark spot exists at high latitudes on the protostar, it is possible that the dark spot affects the inner disk except for the innermost part of the disk, which could simultaneously explain why the shape of CO overtone bandhead and of the highest velocity components of the atomic lines does not change. We were unable to note a variation in the CO overtone bandtail (see lower middle panel in Figure~\ref{fig:co_obs}) because of the low signal to noise ratio and poor telluric correction.

In both sources, the observed blue/red peaks  are at $\sim$50--60 km~s$^{-1}$ from line center, which are similar to the velocities at the outer radius in the best-fit models for the CO overtone band emission.
We compared the Na I and Ca I spectral features with a simple Keplerian disk emission model.
For this comparison, we adopted a power-law distribution of the normalized intensity ($I(r'_{\rm in})=1$) with a power-law index of $p'$.
This model produces the emission profile peaked around the rotation velocity of the outermost annulus ($r'_{\rm out}$) while the emission at the highest velocities is radiated from the innermost annulus \citep[$r'_{\rm in}$,][]{Smak1981}. Low signal-to-noise ratios make it difficult to constrain the highest velocity from our observed line profiles.
 We therefore set the inner radius to be the same as in the best-fit CO band emission models, i.e., $r'_{\rm in}=r_{\rm in}$. The value of $\Delta V_{\rm FWHM}$ is also adopted from the CO band emission models. Therefore, the free model parameters for the Na I and Ca I spectra are $r'_{\rm out}$ and $p'$ only. The measured velocity ($\sim$50--60 km~s$^{-1}$) of the emission peaks constrains the outer radius to 2 $r'_{\rm in}$.
 
The cyan lines in Figures~\ref{fig:atom-i03} and \ref{fig:atom-i04} represent the best-fit model spectra with $r'_{\rm out}/r'_{\rm in}\sim 2$ and $p' \sim 0.0$ for both sources. Flux peaks of the synthesized spectra are adjusted to fit the highest peaks in the most recent observations (red lines). The Na I and Ca I spectra cannot be explained solely by a simple axisymmetric Keplerian disk, though this model fits the observed spectra on 19 Nov. 2015 for IRAS~04239+2346 rather well.

\subsection{The warm gas at larger radii} \label{ss:out}
\begin{figure*}
\includegraphics[width=0.31 \textwidth]{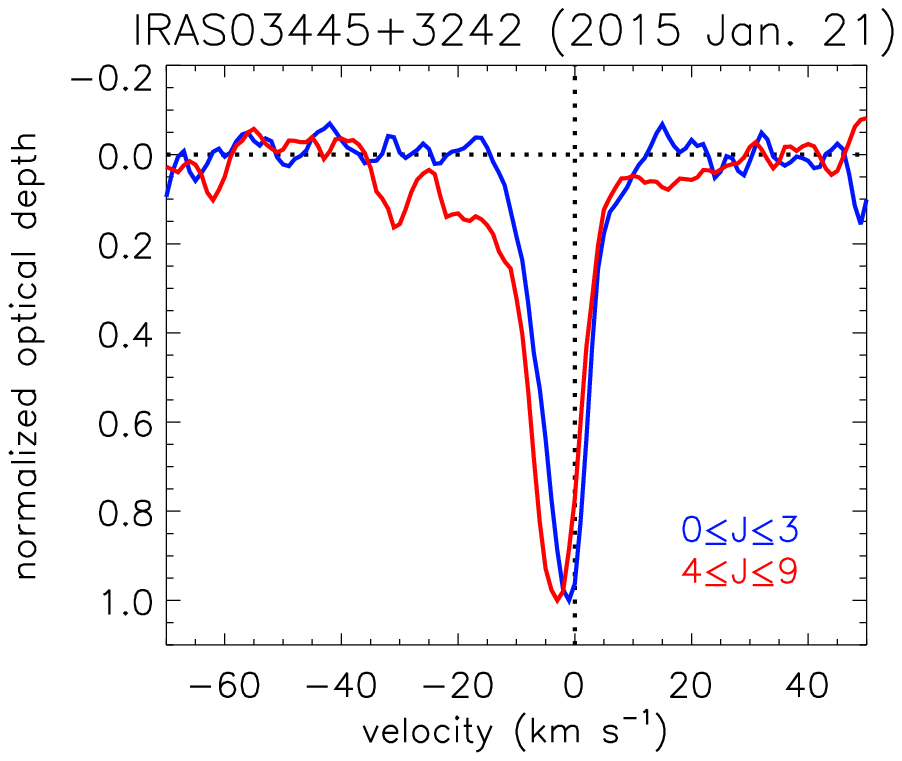}
\includegraphics[width=0.31 \textwidth]{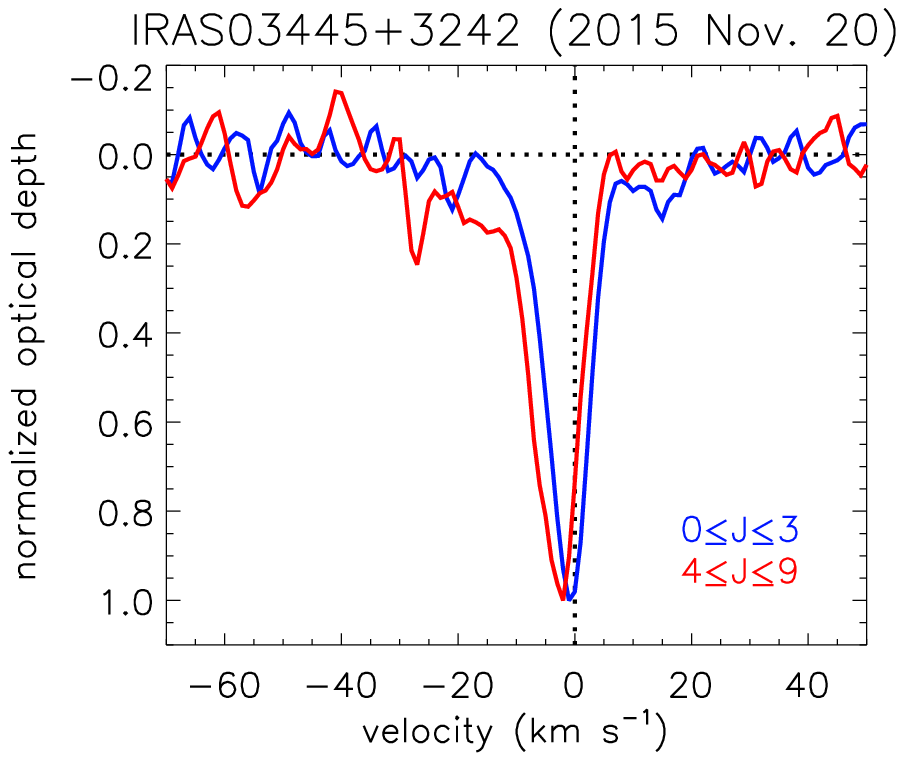}

\caption{Averaged line profiles of lower-$J$ ($0\le J\le 3$ for blue spectra) and higher-$J$ ($4\le J\le 9$ for red spectra) CO transitions for IRAS~03445+3242. The optical depth is normalized to the absorption peak. }
\label{fig:co_line_profile}
\end{figure*}

\begin{figure*}
\includegraphics[width=0.31 \textwidth]{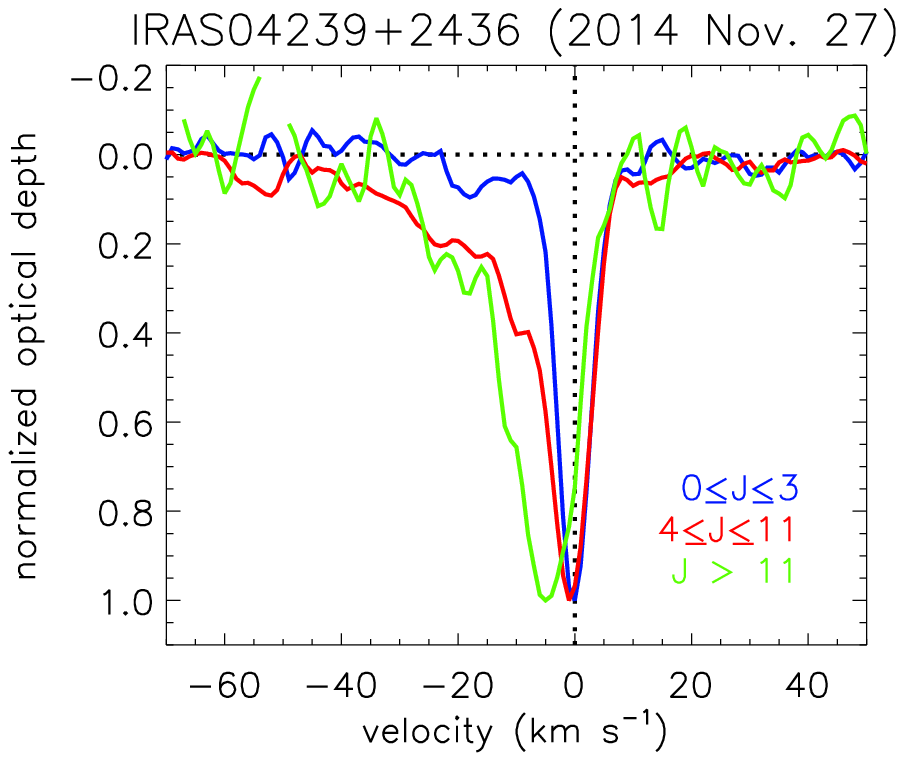}
\includegraphics[width=0.31 \textwidth]{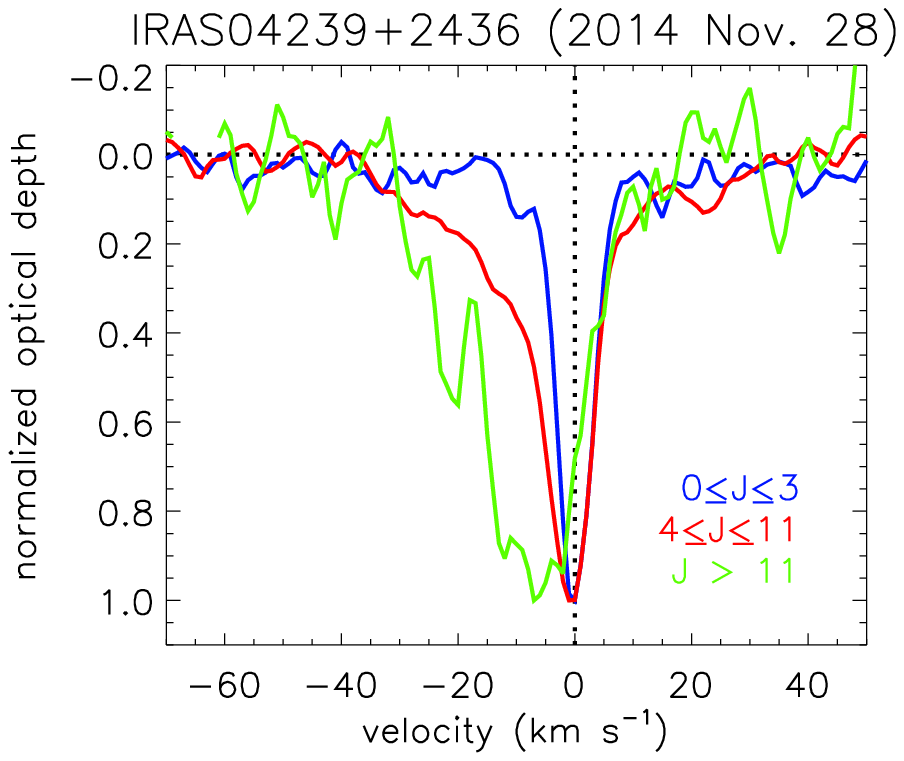}
\includegraphics[width=0.31 \textwidth]{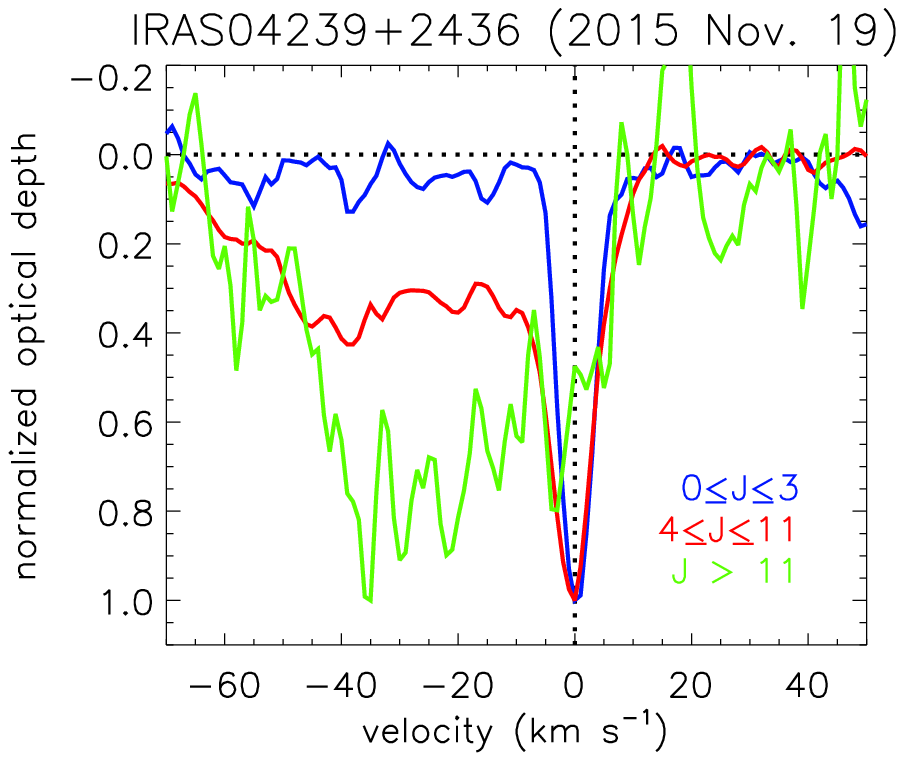}
\caption{Averaged line profiles of lower-$J$ ($0\le J\le 3$ for blue spectra) and higher-$J$ ($4\le J\le 11$ for red spectra and $J>$ 11 for green spectra) CO transitions for IRAS~04239+2436. The optical depth is normalized to the absorption peak.}
\label{fig:co_line_profile1}
\end{figure*}

\begin{figure*}

\includegraphics[width=0.31 \textwidth]{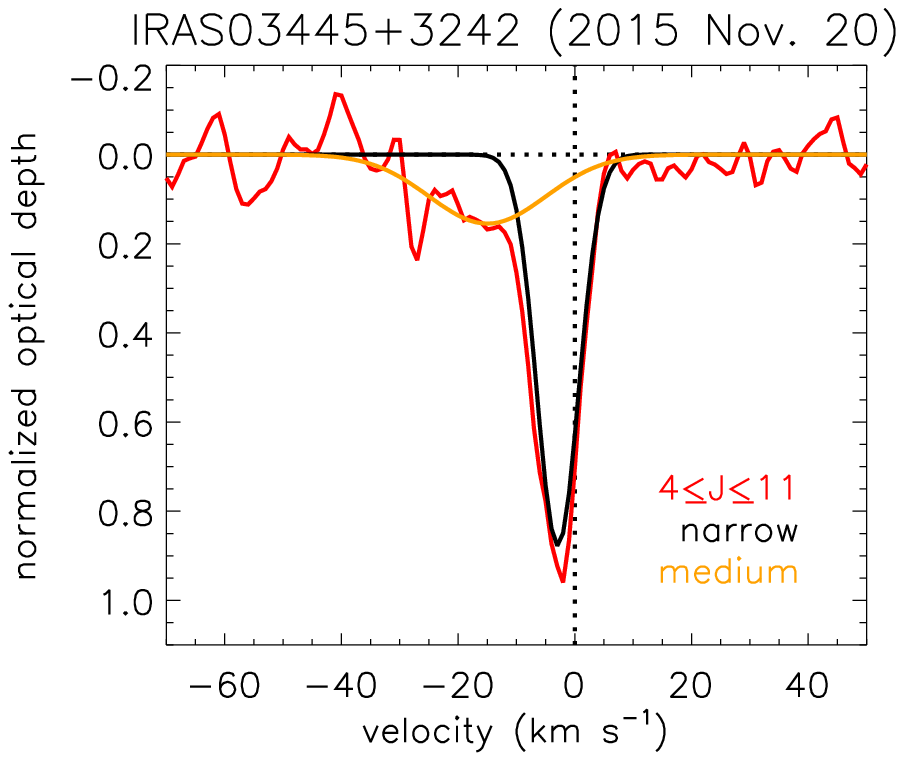}
\includegraphics[width=0.31 \textwidth]{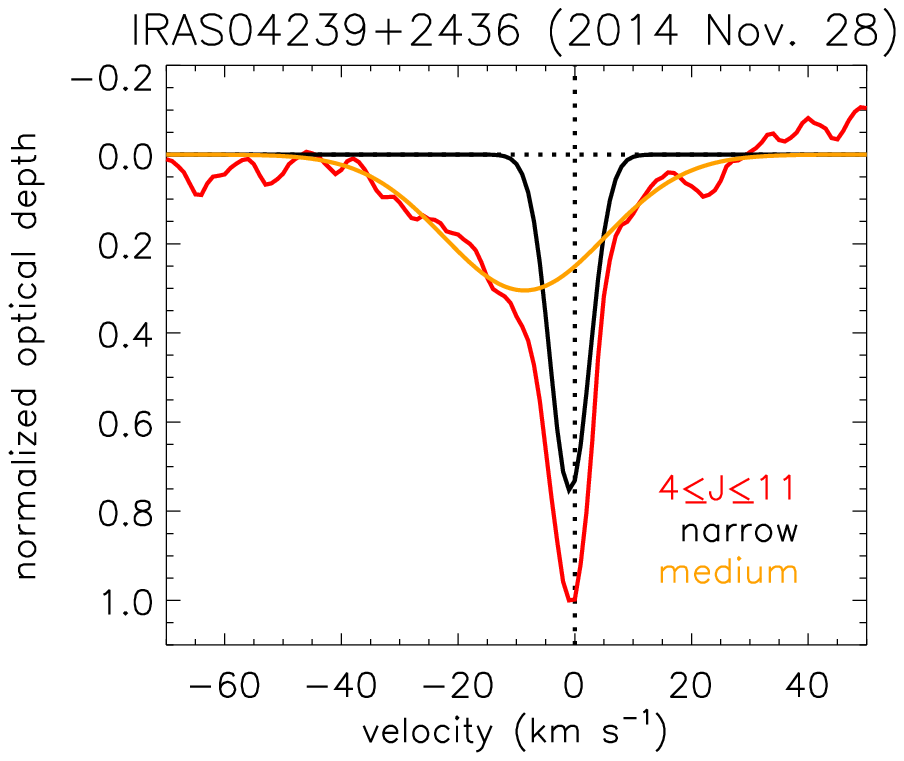}
\includegraphics[width=0.31 \textwidth]{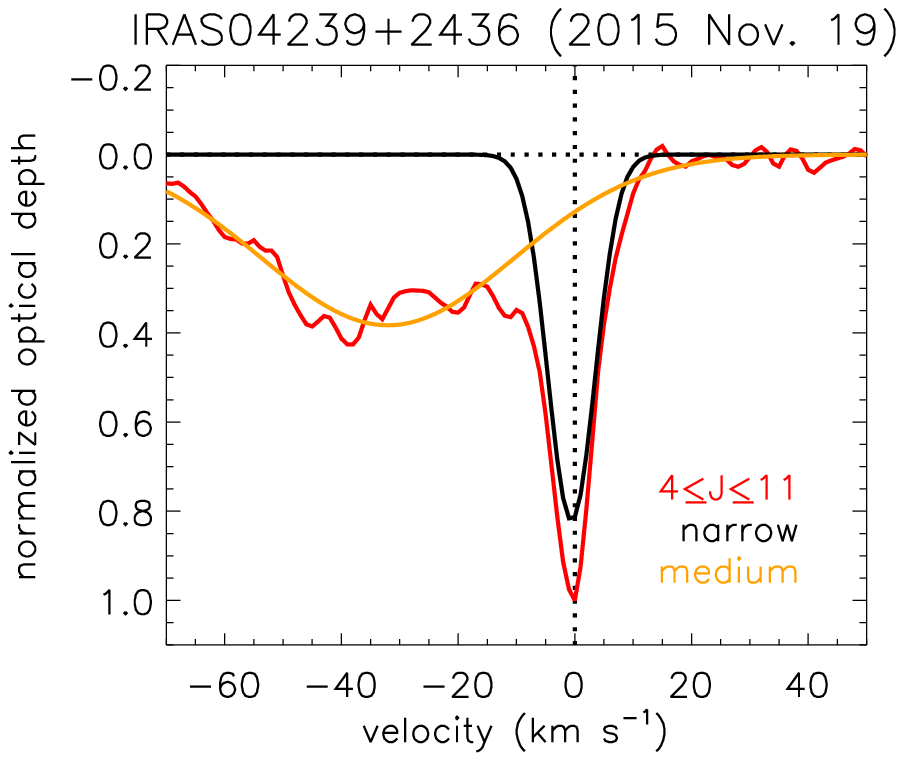}
\caption{Gaussian decomposition of the higher-$J$ ($4\le J\le 9$) CO transitions for IRAS~03445+3242 (left) and the higher-$J$ ($4\le J\le 11$) ones for IRAS~4239+2436 observed in the second (middle) and the last (right) epochs: a deep spectrally unresolved component ($\sim$8~km~s$^{-1}$; black line) and a shallow medium-width component ($\sim$30-50~km~s$^{-1}$; orange line). The first epoch observation has a similar decomposition to the second epoch observation for both sources. The medium-width component in IRAS~03445+3242 is too weak to be analyzed quantitatively. However, in IRAS~04239+2436, the medium-width component is clearly detected, and its central velocity and line with become more blueshifted and broader in one year.}
\label{fig:co_line_profile2}
\end{figure*}

The right panels of Figure~\ref{fig:co_obs} show that there are narrow absorption lines superposed on the emission in the CO overtone bands. These absorption lines are the lower-$J$ rovibrational transitions of the first overtone band ($v=0\rightarrow2$). In considering and modeling these absorption features, we make the reasonable assumption that the continuum and the broad CO overtone bands lie behind the absorbing medium along the line of sight. Averaged absorption line profiles toward IRAS~03445+3242 and IRAS~04239+2436 are plotted in Figure~\ref{fig:co_line_profile} and \ref{fig:co_line_profile1}, respectively. Higher-$J$ absorption line profiles  in IRAS~04239+2436 seem to be composed of deep \textit{spectrally unresolved} ($FWHM\sim$8~km~s$^{-1}$) and shallow \textit{medium-width} ($FWHM\sim$30-50~km~s$^{-1}$) components as fit in Figure~\ref{fig:co_line_profile2}. The shallow medium-width component is also present in the higher-$J$ absorption profiles toward IRAS~03445+3242, although it is very weak. Therefore, the absorption lines in both sources were fit by multi-Gaussian profiles (with a 1st order polynomial fit to the base line). We first analyze the \textit{spectrally unresolved} component.

\begin{figure*}

\includegraphics[width=0.45 \textwidth]{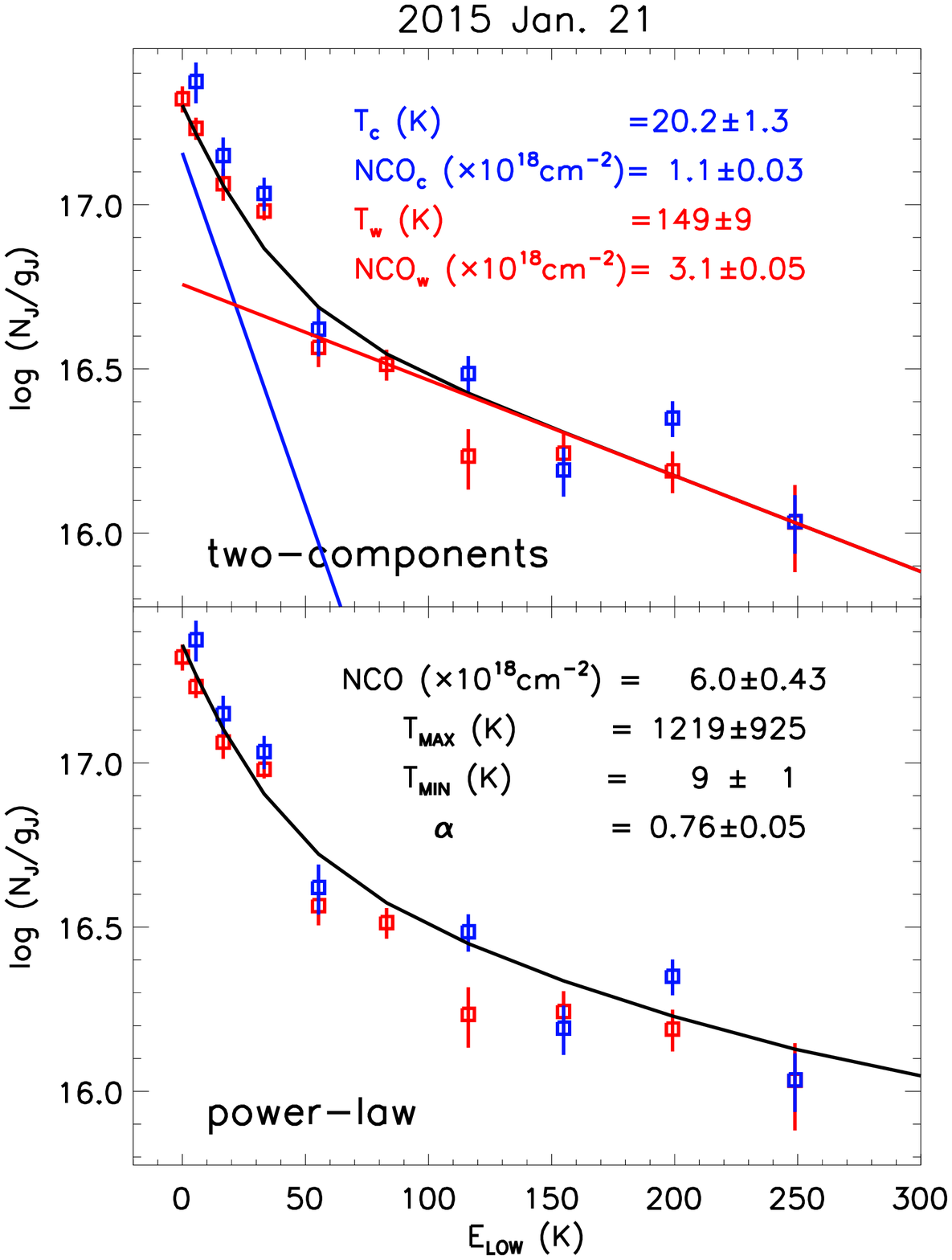}
\includegraphics[width=0.45 \textwidth]{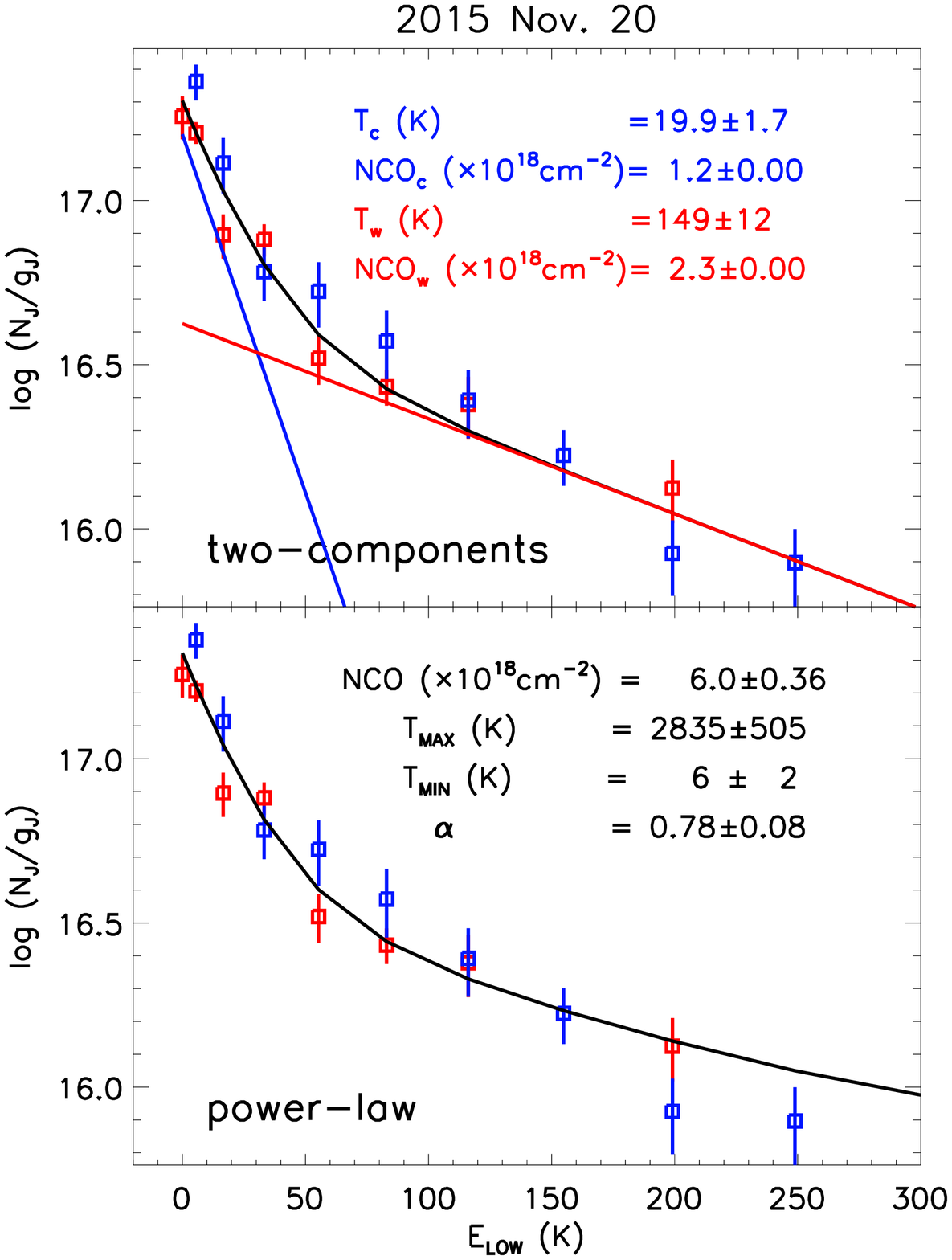}
\caption{Rotation diagrams of the \textit{spectrally unresolved} CO rovibrational absorption spectra ($v=0\rightarrow2$) for IRAS~03445+3242. The squares represent the observed column densities of R- (red squares) and P-branch (blue squares) transitions, divided by statistical weights ($N_J/g_J$).  The vertical bars indicate the 1 $\sigma$ error. Top : the rotation diagram is fit with two distinct cold (blue line) and warm (red line) components. The column density and gas temperature of each component for the best-fit model are presented as the same colors on the panel. The black solid line indicates the sum of the two components. Bottom: the rotation diagram is fit with a power-law model ($dN \propto T^{-\alpha} dT$). The column density, the maximum and minimum temperatures, and the power index of the best-fit model are presented in the panel. 
}
\label{fig:abs-i03-0}
\end{figure*}

\begin{figure*}
\includegraphics[width=0.32 \textwidth]{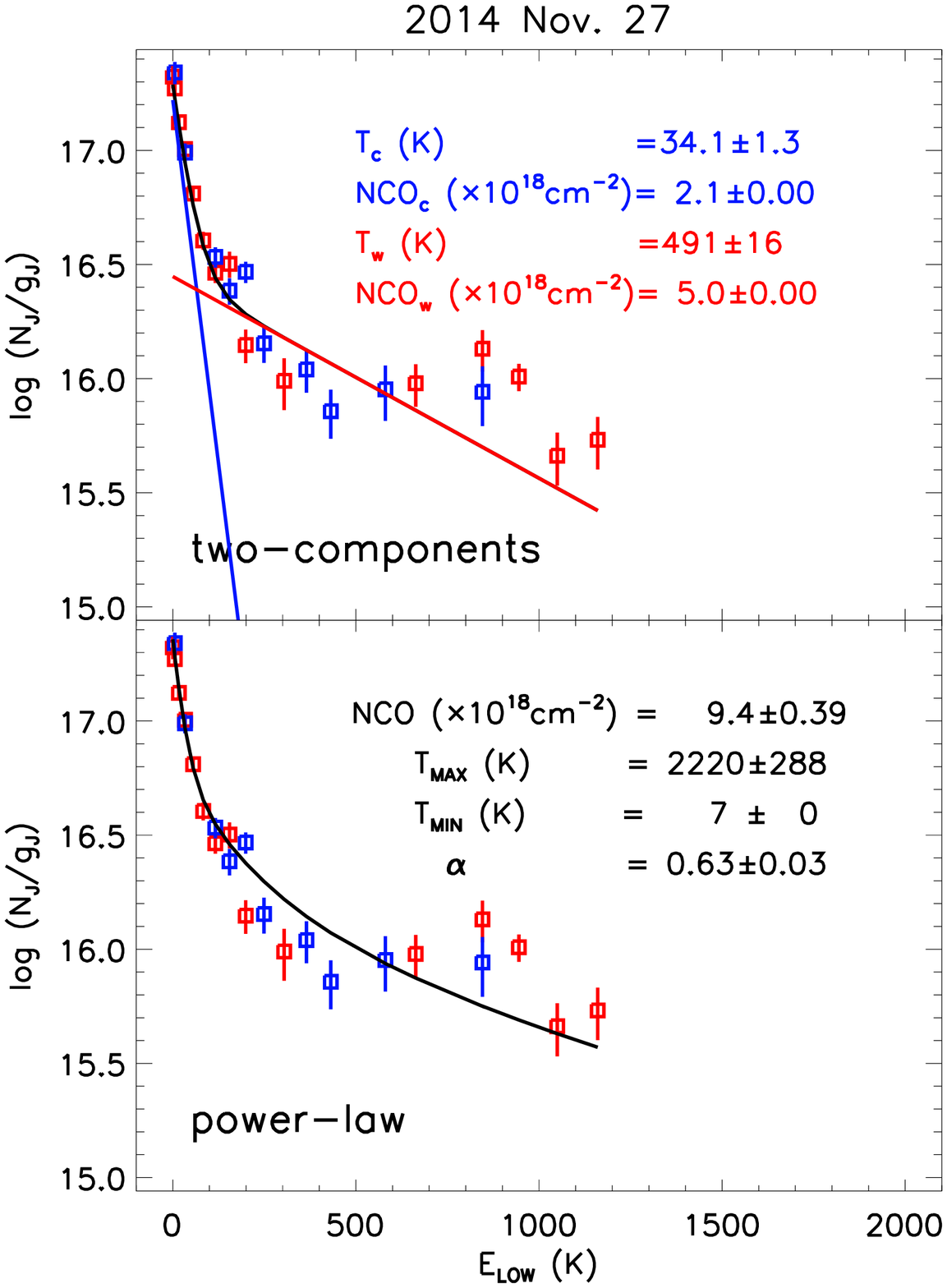}
\includegraphics[width=0.32 \textwidth]{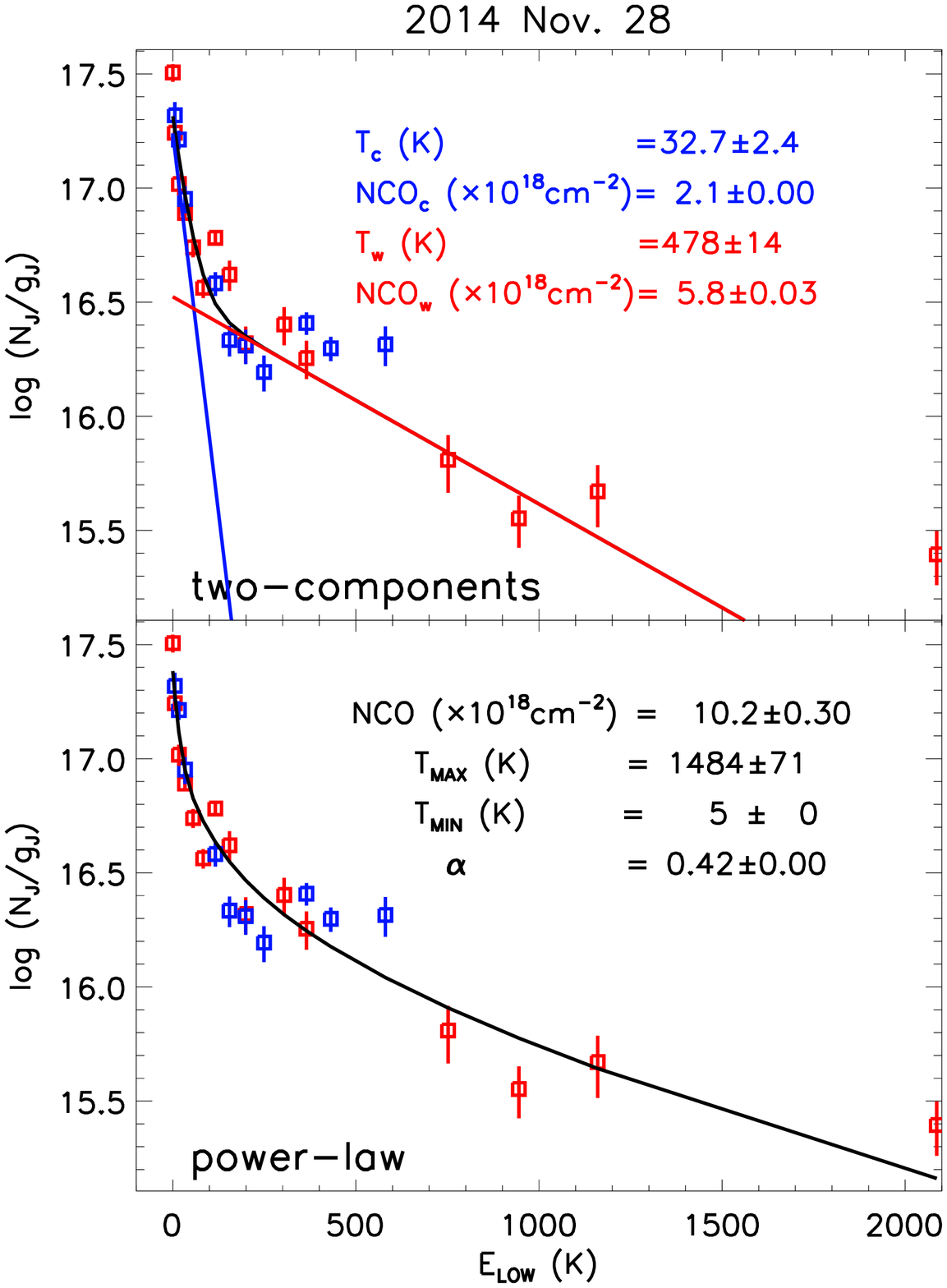}
\includegraphics[width=0.32 \textwidth]{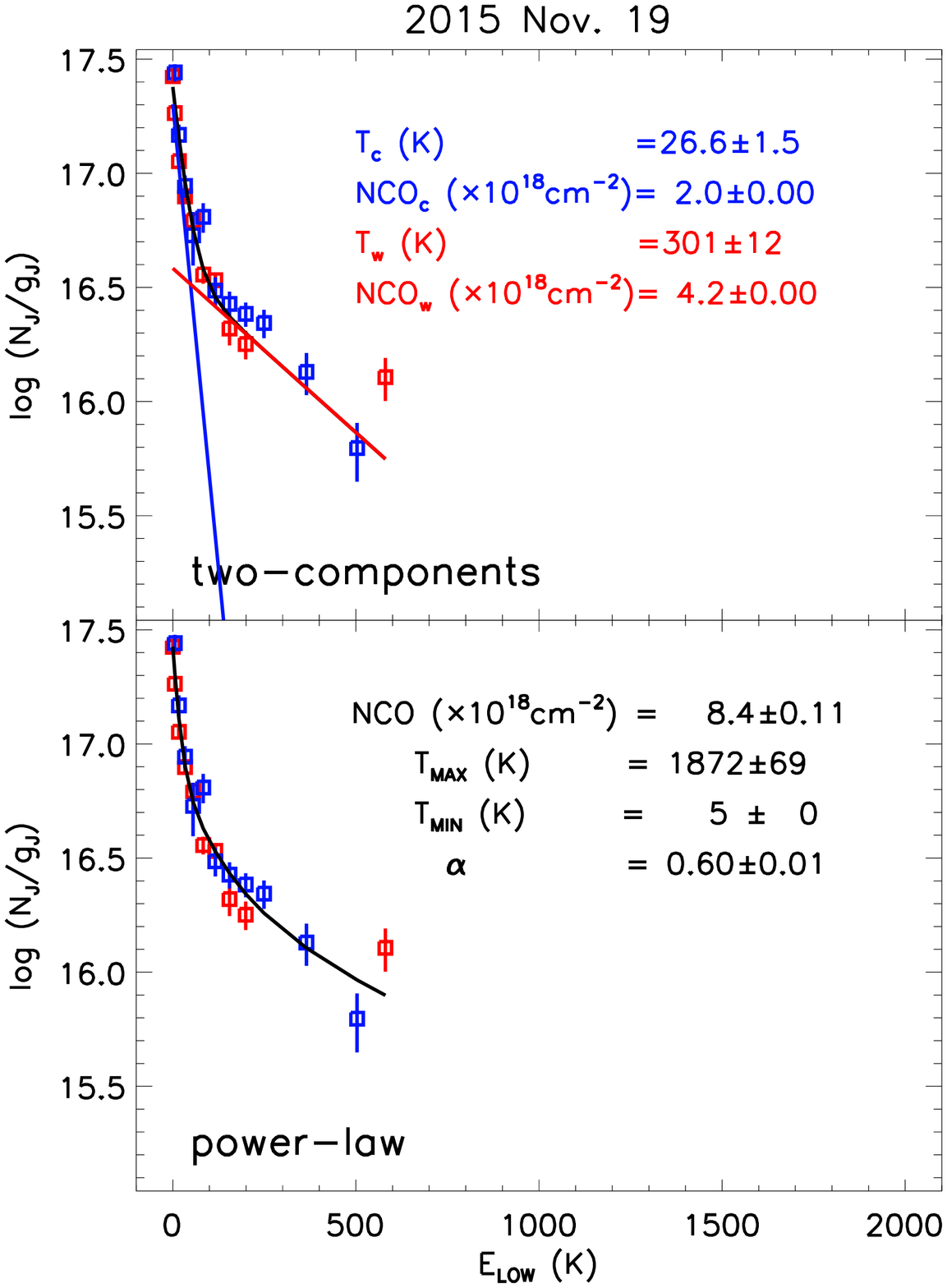}
\caption{Same as Figure~\ref{fig:abs-i03-0} except for IRAS~04239+2436.}
\label{fig:abs-i04-0}
\end{figure*}

 Assuming optically thin absorption, we calculated the column density for each rovibrational absorption transition; the column densities are divided by the statistical weights to make an excitation diagram (Figures~\ref{fig:abs-i03-0} and \ref{fig:abs-i04-0}). Although both the R and P branches should have the same column density, in some cases, only one branch is detected or a difference between the column densities derived from two branches is larger than the 3 $\sigma$ error. This inconsistency may be caused by imperfect telluric correction (see below).
 
 The spectrally unresolved CO rovibrational absorption transitions show positive curvature in the excitation diagrams for both targets, indicative of multiple components.
 Therefore, we tried to fit the rotation diagrams with two distinct temperature components: free parameters are a warm gas temperature ($T_{\rm w}$), a cold gas temperature ($T_{\rm c}$), and the ratio of the column density of warm CO gas to that of cold CO gas.  The column density of the cold CO gas, $N({\rm CO})_{\rm c}$, is determined by minimization of $\chi^2$ in fitting the observations ($\partial \chi^2 / \partial N(\rm{CO})_{\rm c} = 0$). 
 
 The parameters for the best-fit models are presented in the top panels of Figures~\ref{fig:abs-i03-0} and \ref{fig:abs-i04-0}. When only thermal line broadening is considered, the best-fit model has a maximum optical depth of $\sim$8 at the $R(2)$ transition, and the optically thin assumption could break down in the lower levels if non-thermal line widths or velocity gradients along the line of sight are too small. Therefore, the column densities of CO in the lower-$J$ levels could be underestimated more significantly than those in the higher-$J$ levels, and our column densities derived from the rotation diagrams may be underestimates while our gas temperature may be overestimates.
 
 The gas temperature of the warm component is $\sim$150~K for IRAS~03445+3242 and 300-490~K for IRAS~04239+2436, respectively. Using the gas temperature distribution from the best-fit model for the CO overtone band emission, the warm gas, which produces the absorption lines, is located at 9.4~AU in IRAS~03445+3242 and 1.2-4.4~AU in IRAS~04239+2436. (If the absorption occurs in the envelope, the warm gas should be present in the innermost part of the envelope.) 

\begin{figure*}
\includegraphics[width=0.45 \textwidth]{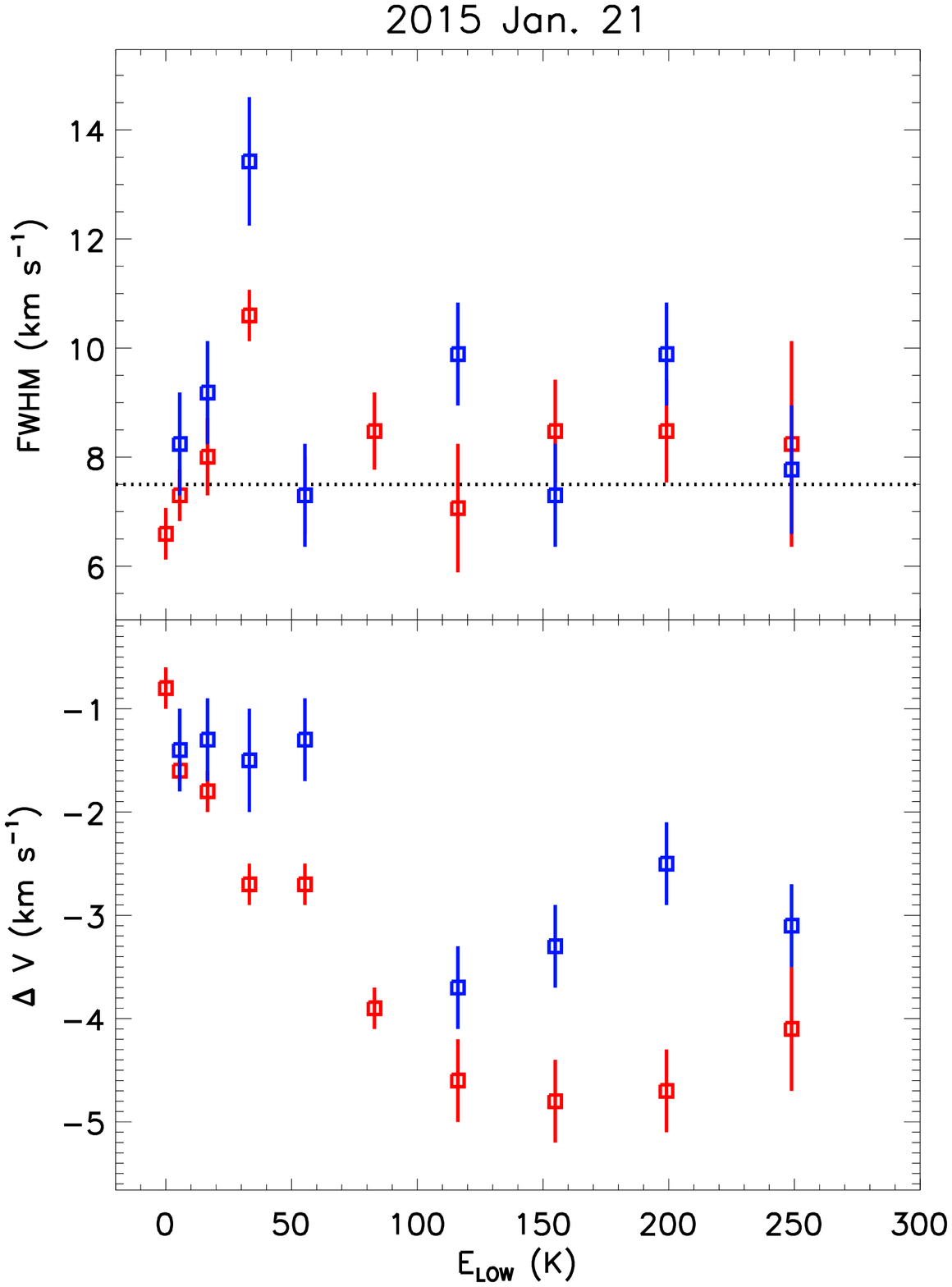}
\includegraphics[width=0.45 \textwidth]{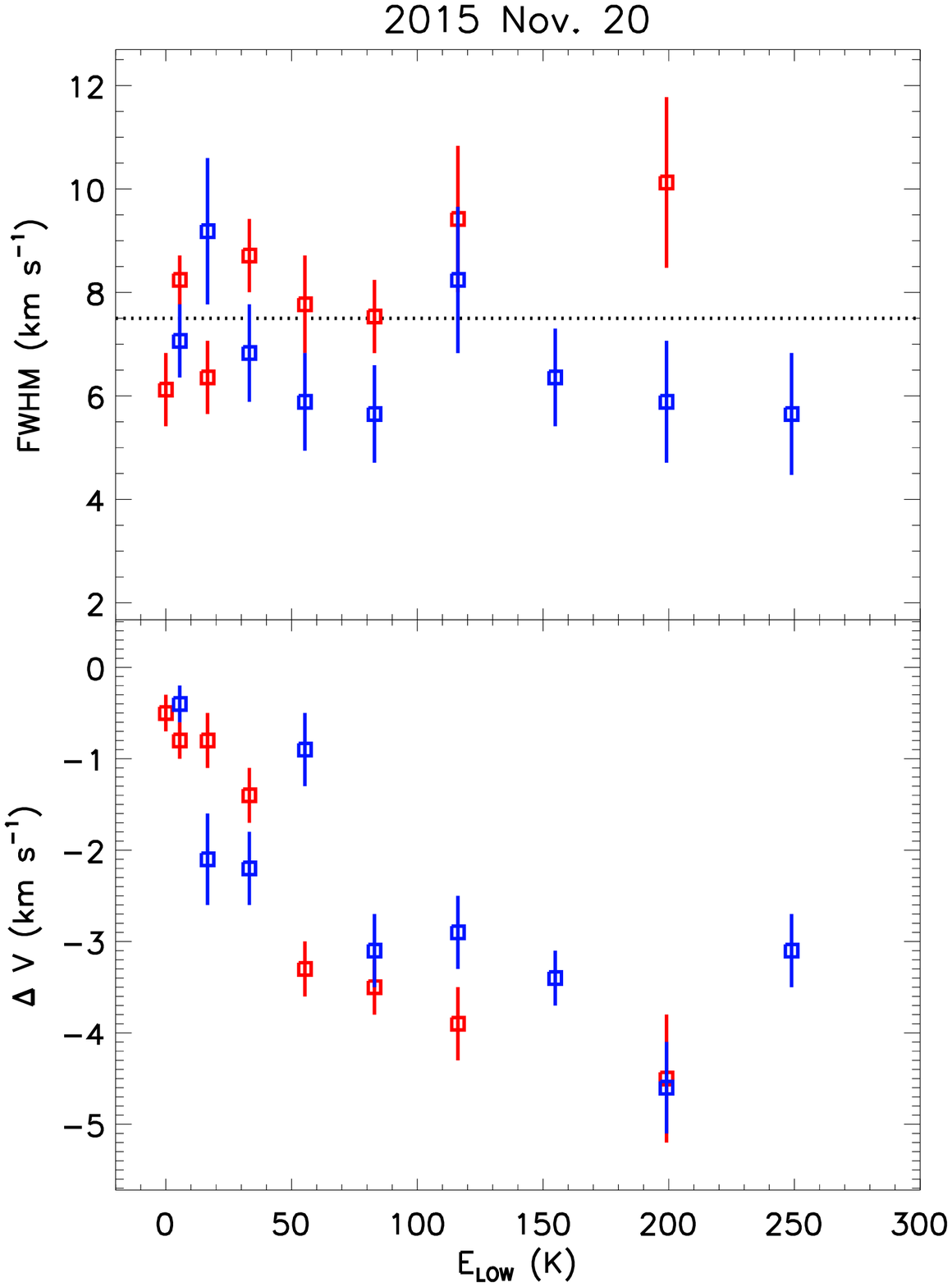}
\caption{Line width (top) and central velocity relative to the source velocity (bottom) of the spectrally unresolved CO rovibrational absorption spectra ($v=0\rightarrow2$) for IRAS~03445+3242.  The line widths are not deconvolved  and thus are influenced by an instrumental velocity resolution of 7.5~km~s$^{-1}$ (presented as horizontal dotted lines in the right top panel). The red and blue squares represent the line widths (top) and the central velocities (bottom) of R- and P-branch transitions, respectively.  The vertical bars indicate the 1$\sigma$ error, which does not take into account the uncertainty in the wavelength calibration ($<\sim$1~km~s$^{-1}$).}
\label{fig:abs-i03-1}
\end{figure*}

\begin{figure*}

\includegraphics[width=0.32 \textwidth]{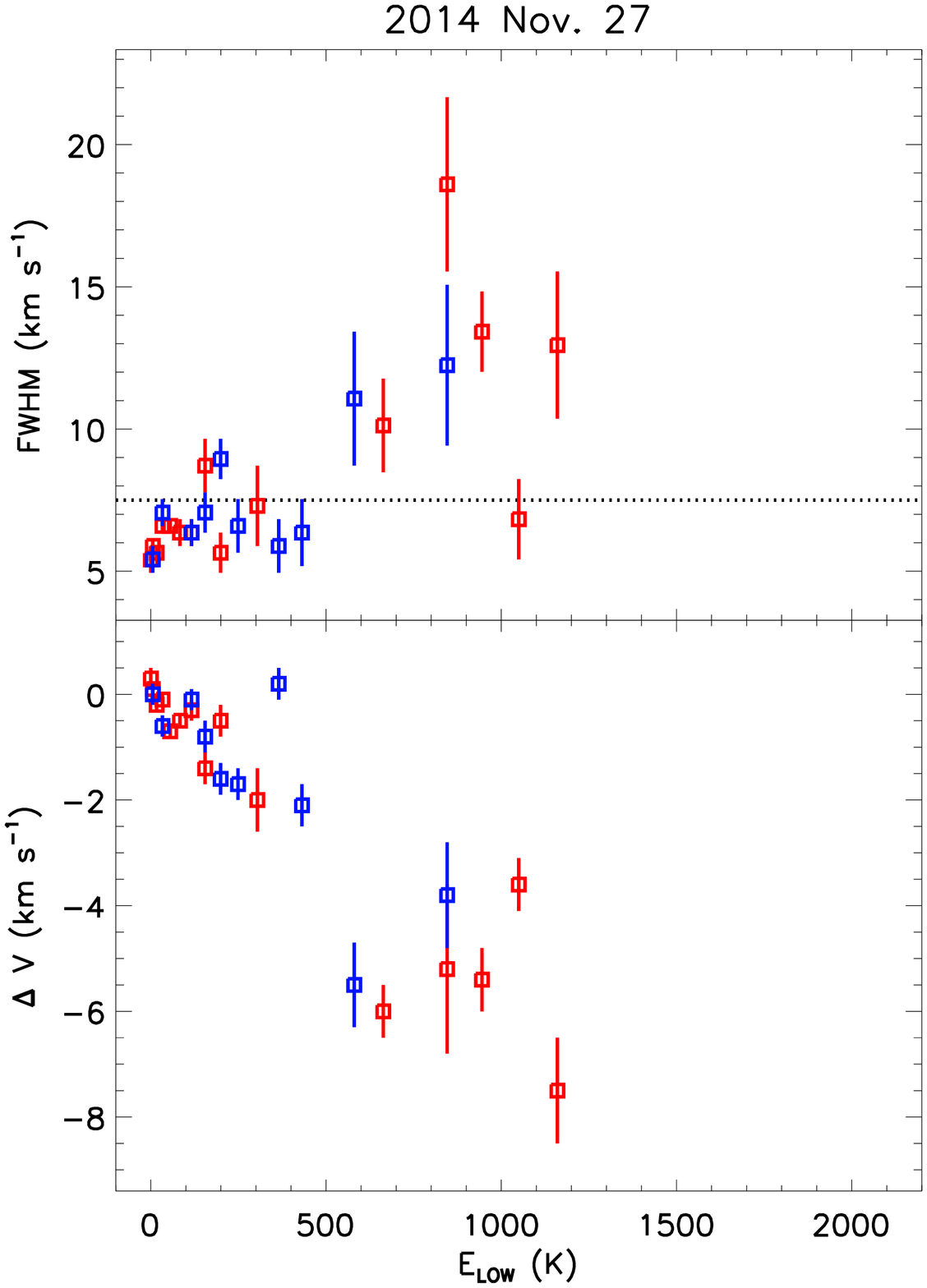}
\includegraphics[width=0.32 \textwidth]{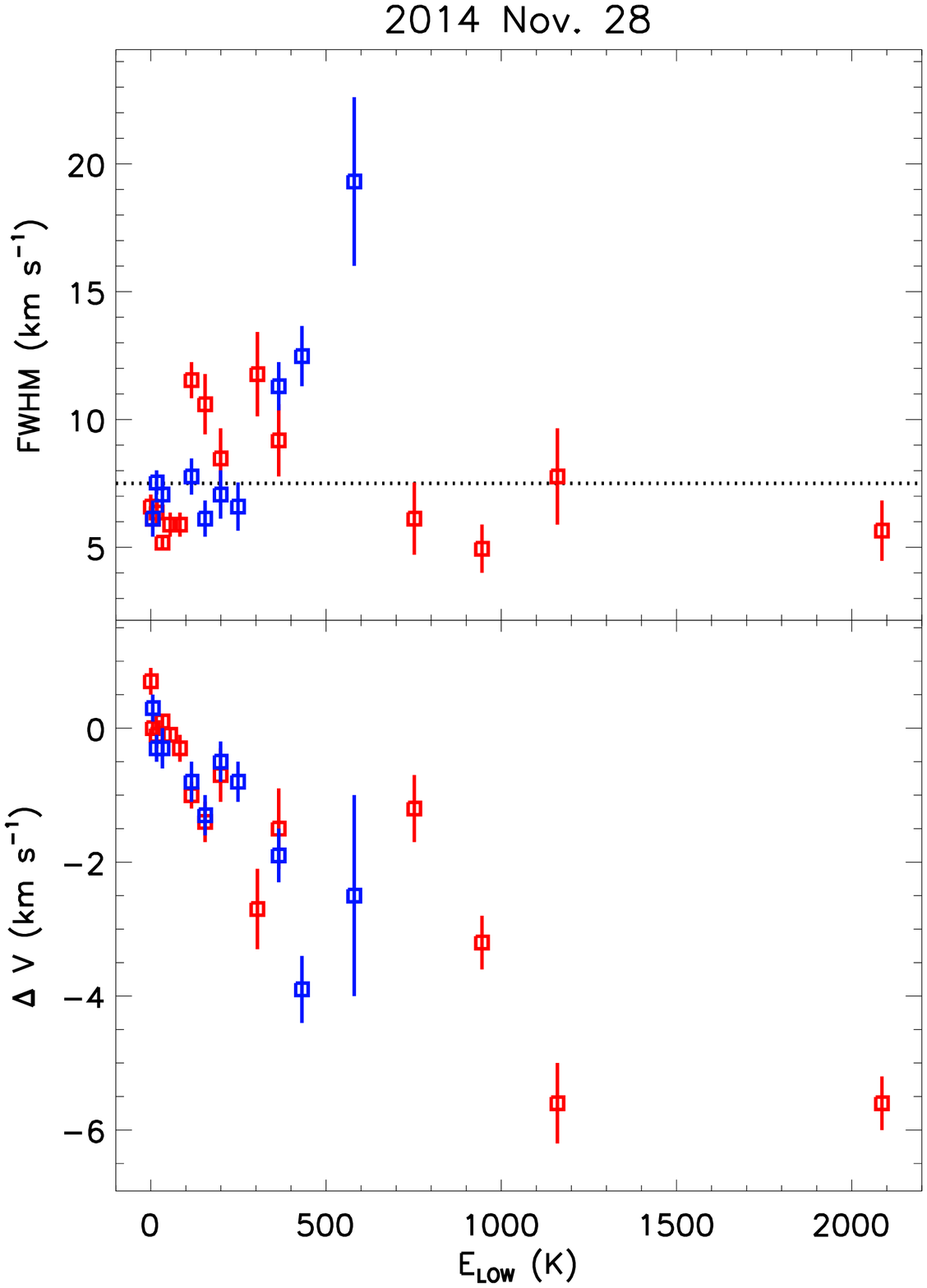}
\includegraphics[width=0.32 \textwidth]{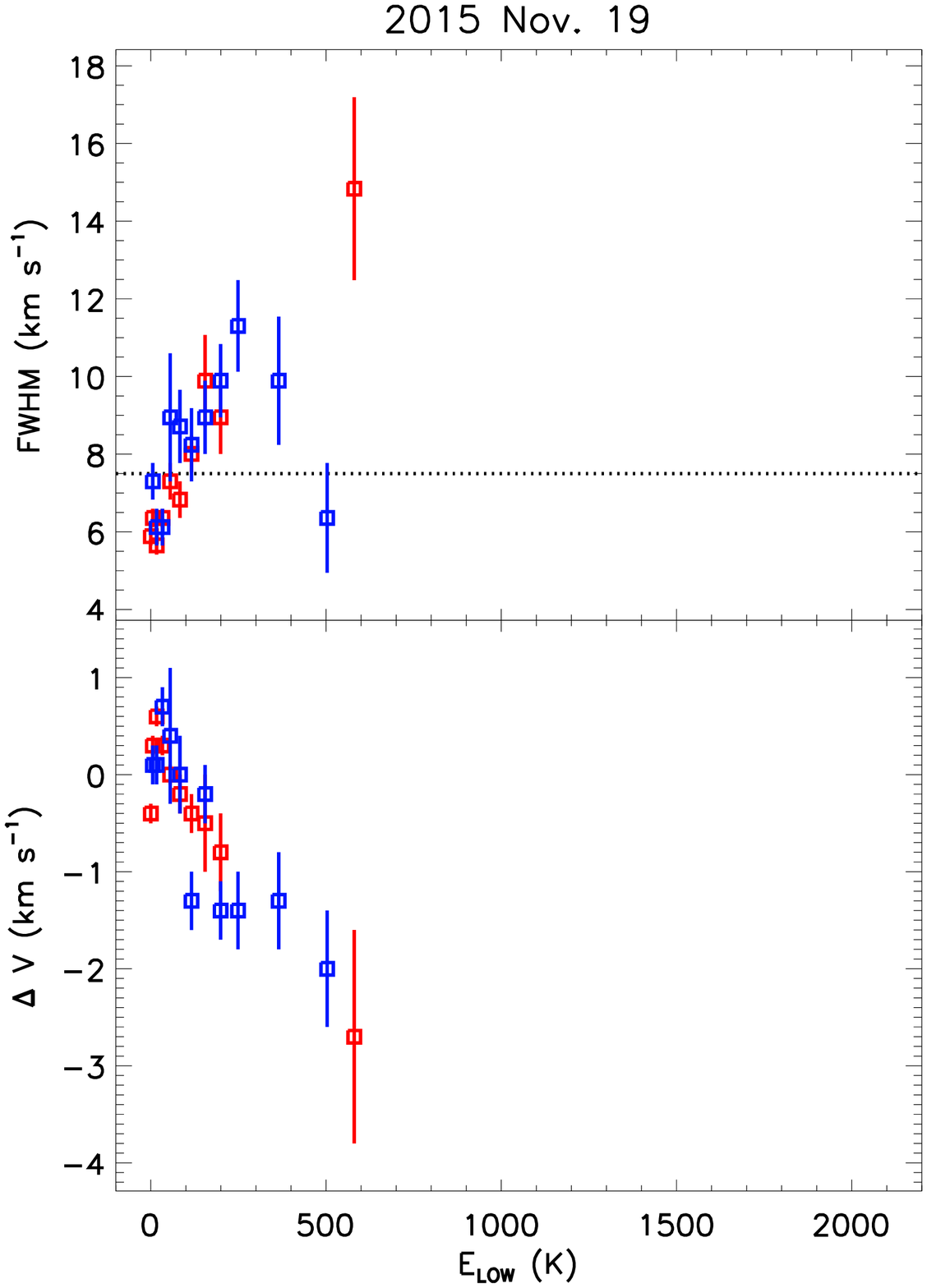}
\caption{Same as Figure~\ref{fig:abs-i03-1} except for IRAS~04239+2436.}
\label{fig:abs-i04-1}
\end{figure*}

 Central velocities (relative to the source velocity) and line widths for the spectrally unresolved CO absorption lines are plotted in Figures~\ref{fig:abs-i03-1} and \ref{fig:abs-i04-1}.
 As shown on the bottom panels in the figures, the higher-$J$ CO lines show a velocity shift ($\sim$5~km~s$^{-1}$) to the blue relative to the source velocity in both sources. 
This is also seen clearly in Figure~\ref{fig:co_line_profile}. For IRAS~03445+3242, the velocity of the averaged spectrum of higher-$J$ lines (red spectrum) is blue-shifted relative to the lower-$J$ lines (blue spectrum) with a similar line width. On the other hand, for IRAS~04239+2436, the averaged spectrum of higher-$J$ lines (4$\le~J\le~11$) shows similar central velocity and line width to those of the lower-$J$ lines ($J<$ 4), but in the average spectrum of much higher transitions ($J> 11$, green lines), the central velocity and the line width became more blueshifted and broader with time (Figure~\ref{fig:co_line_profile1}). 

  The velocities and line widths of corresponding R and P branch lines, which should be identical in principle, show some difference; P branch absorption lines have velocities about $\sim$1~km~s$^{-1}$ less than R branch ones in IRAS~03445+3242 observed on 21 Jan. 2015 (lower left panel in Figure~\ref{fig:abs-i03-1}), which is comparable to the uncertainty of the wavelength solution. However, this phenomenon disappears in the observation one year later (lower right panel in Figure~\ref{fig:abs-i03-1}), while  the overall features such as gas temperatures, CO column densities, and the averaged absorption line shapes do not vary. Therefore, the velocity difference between the R and P  branch lines observed on 21 Jan. 2015 could arise from the small differences of the wavelength solution as well as the contamination by nearby telluric lines.  In addition, the line widths of some transitions are smaller than the width of the instrument profile as shown in the top panels in Figures~\ref{fig:abs-i03-1} and \ref{fig:abs-i04-1} although the line widths of the telluric lines are larger than $\sim$7-8~km~s$^{-1}$. These might be also caused by imperfect telluric correction and wavelength solution and could result in a larger scatter of the rotation diagram (Figures~\ref{fig:abs-i03-0} and \ref{fig:abs-i04-0}).

 Cold ($\sim$20-30~K) gas along the line of sight also contributes to the absorption features. 
 The adopted visual extinction also implies a CO column density of $\sim10^{18}$~cm$^{-2}$ \citep{Pineda2010}, which is similar to our derived CO column density of the cold component. If we consider the warm component, which also contributes to the extinction, the gas to dust ratio is possibly larger than the typical value of 100. This might be indicative of dust stratification in the disk (see \citealt{Rettig2006}). However, the visual extinction is one of the lowest values given in the literature \citep{Davis2011}, so more study is needed.

\begin{figure*}
\includegraphics[width=0.5 \textwidth]{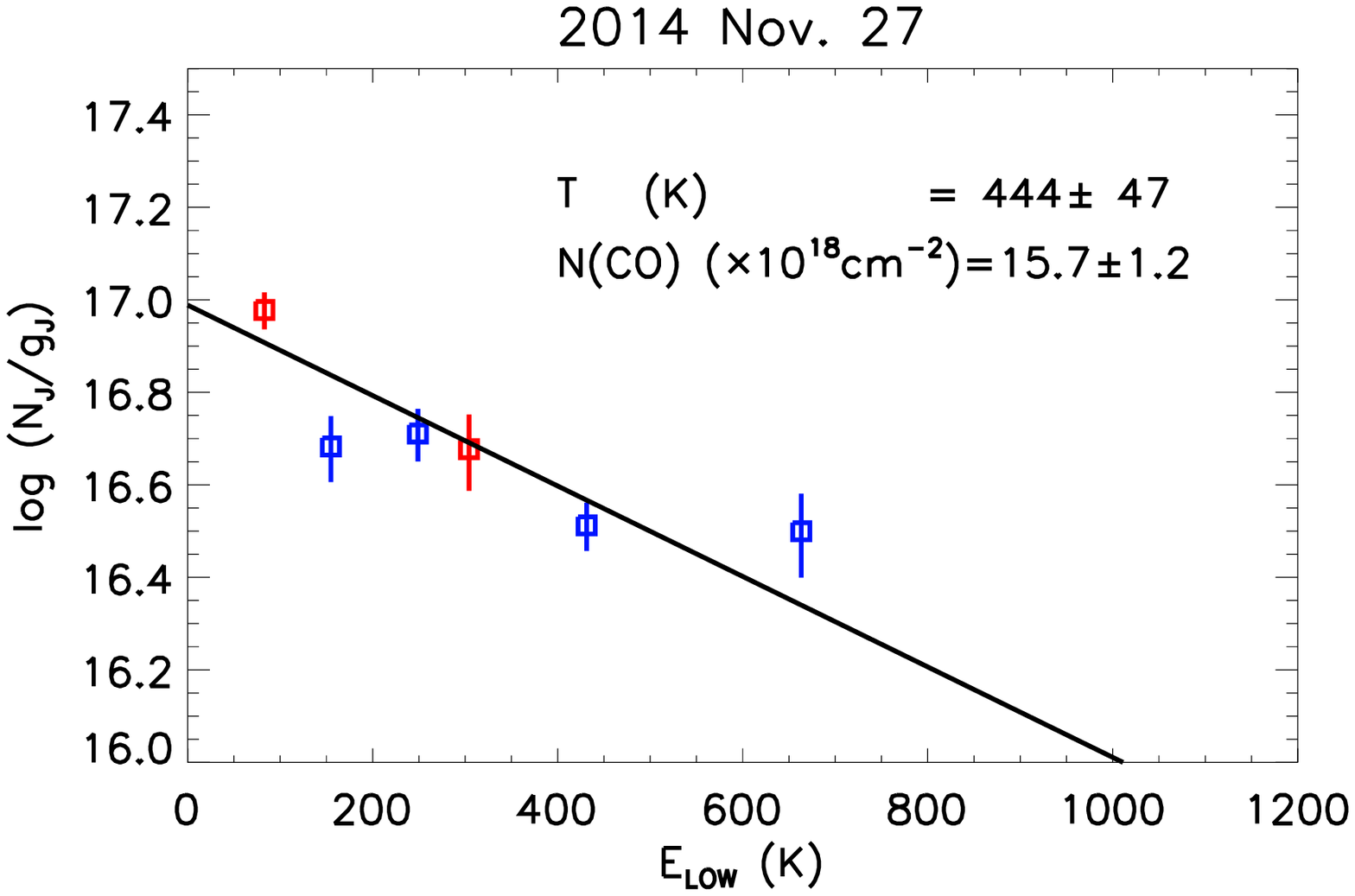}
\includegraphics[width=0.5 \textwidth]{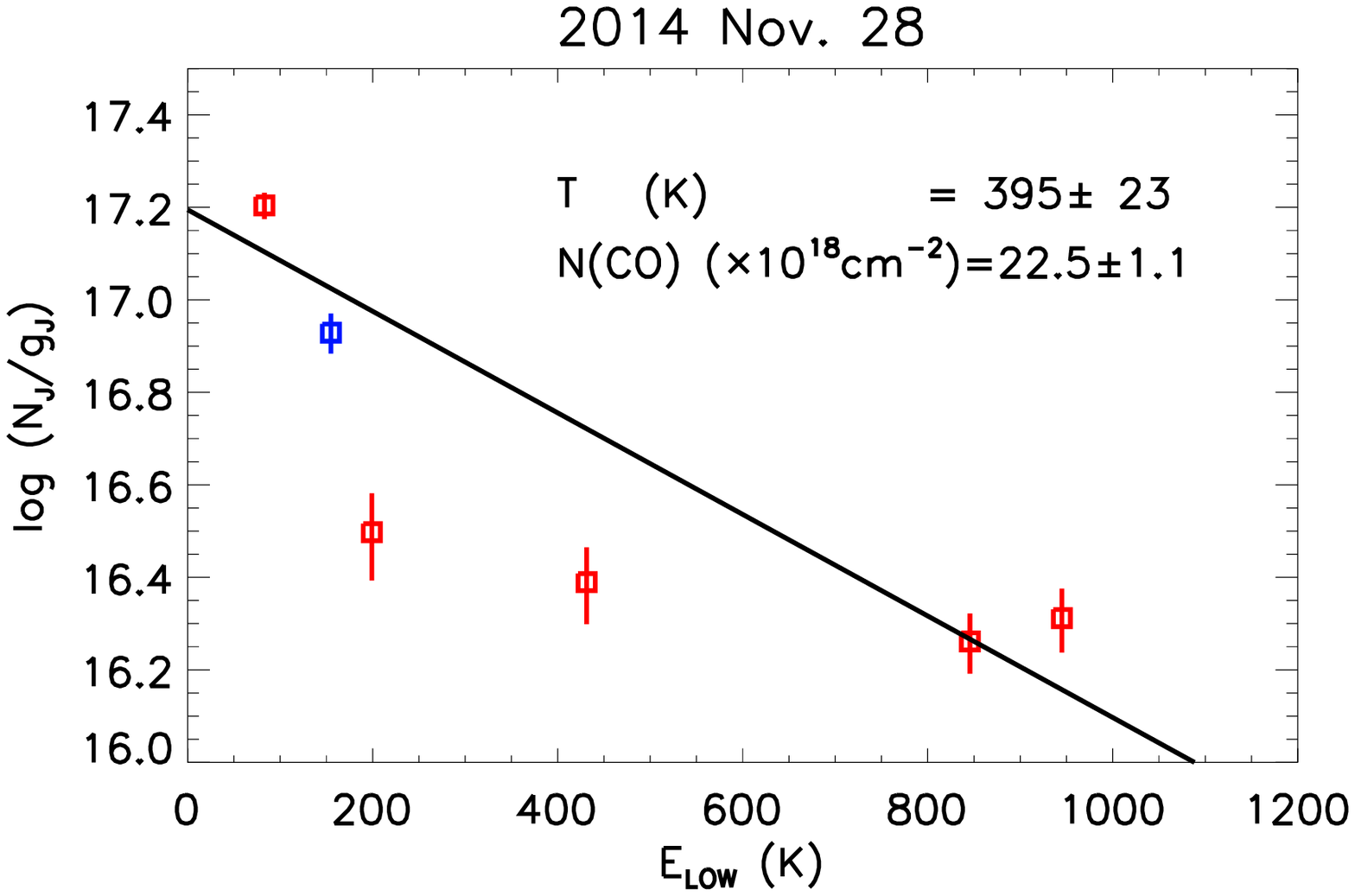}
\includegraphics[width=0.5 \textwidth]{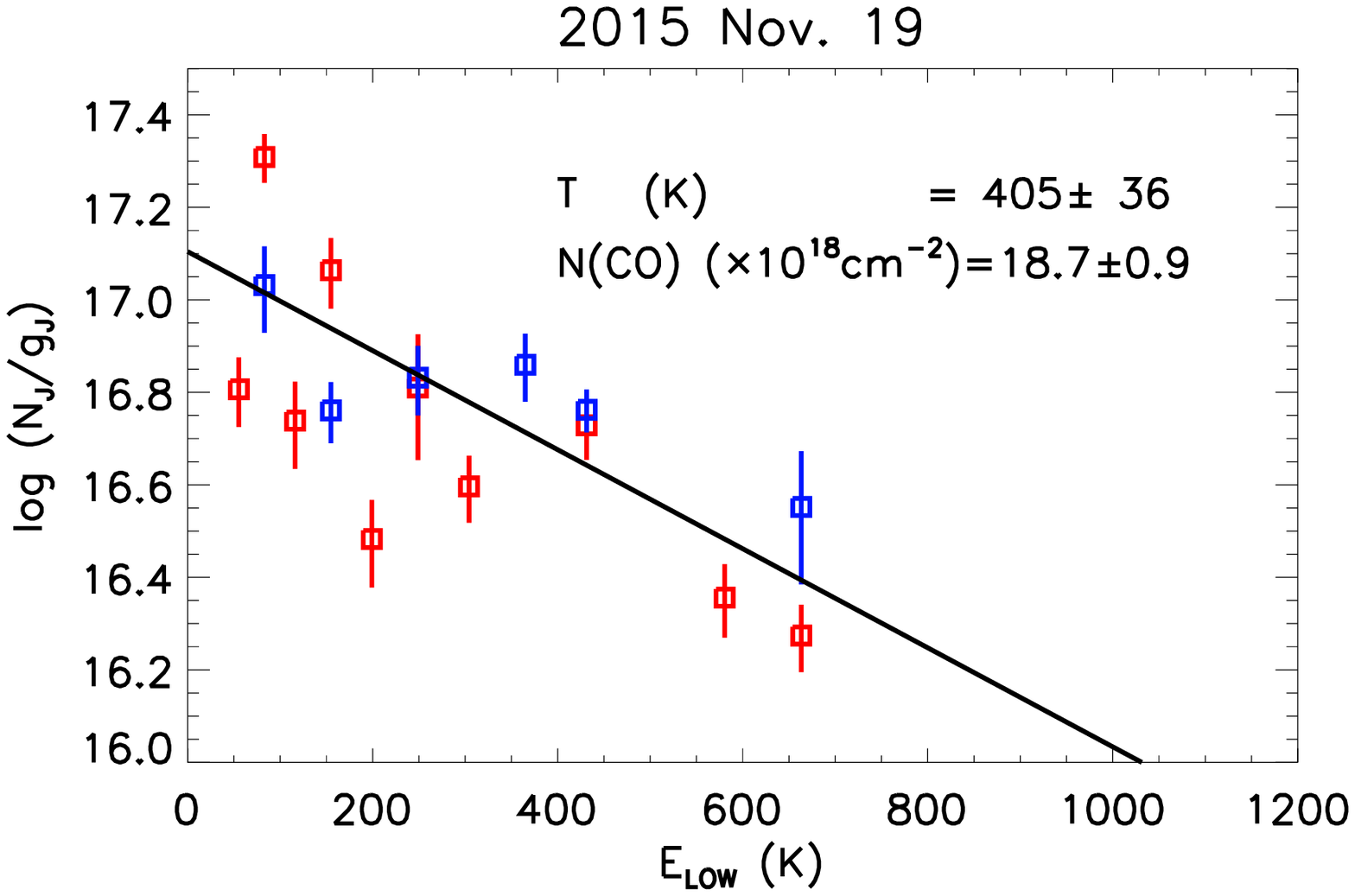}
\caption{Rotation diagrams of the \textit{medium-width} CO rovibrational absorption spectra ($v=0\rightarrow2$) for IRAS~04239+2436. The symbols are the same as those in Figure~\ref{fig:abs-i03-0}. The overplotted black lines correspond to a linear fit to the rotation diagrams.}
\label{fig:abs-i04B}
\end{figure*}

\begin{figure*}
\includegraphics[width=0.32 \textwidth]{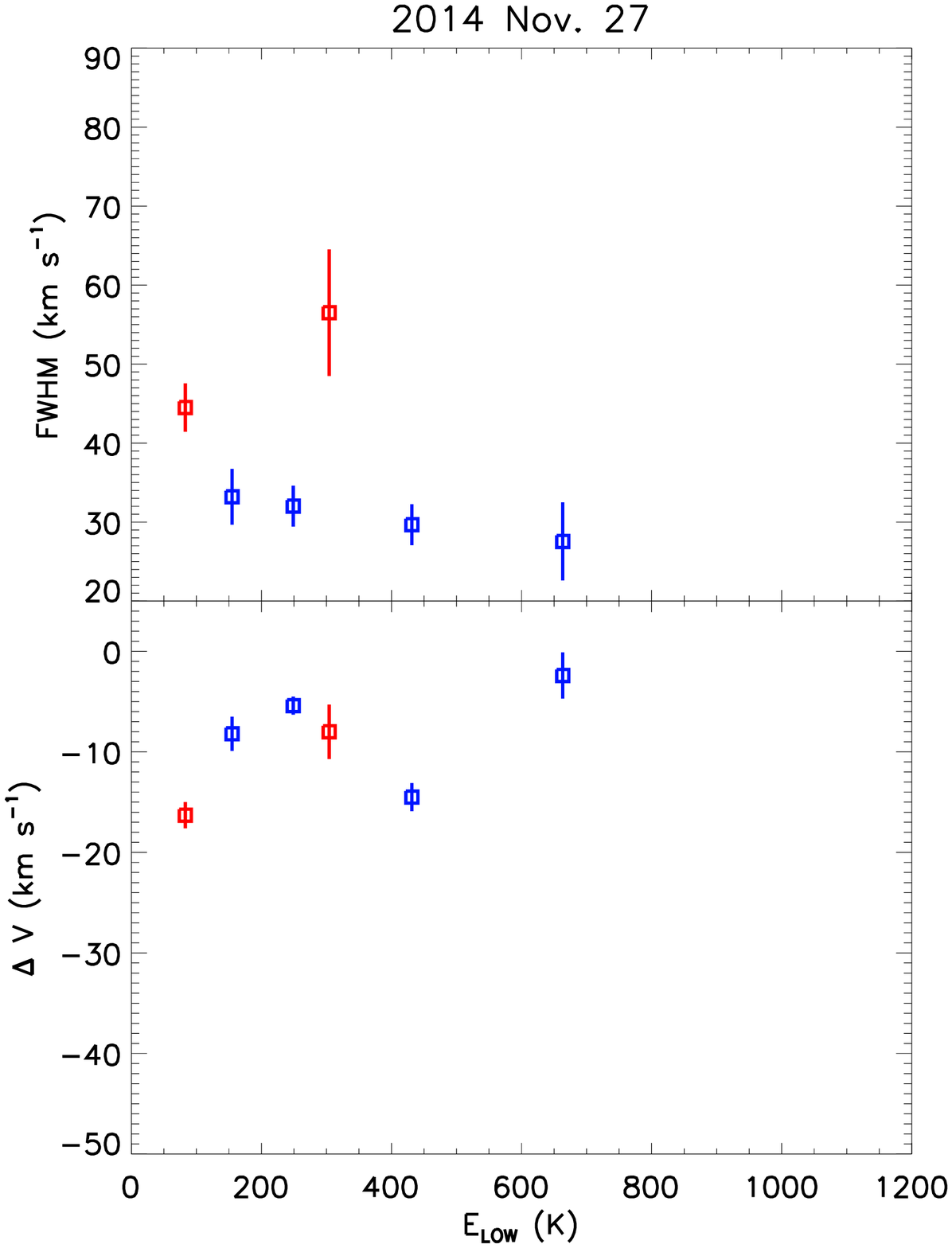}
\includegraphics[width=0.32 \textwidth]{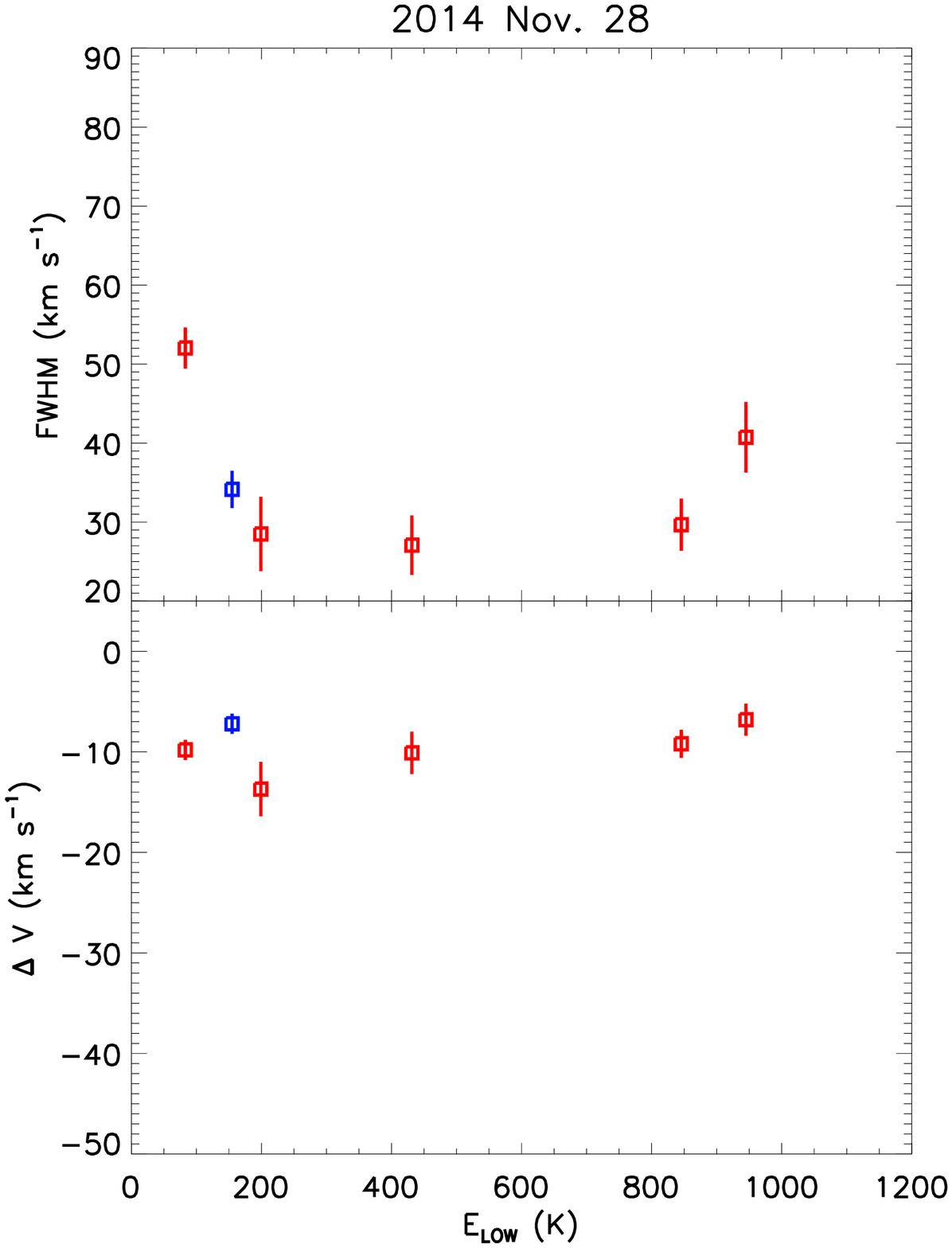}
\includegraphics[width=0.32 \textwidth]{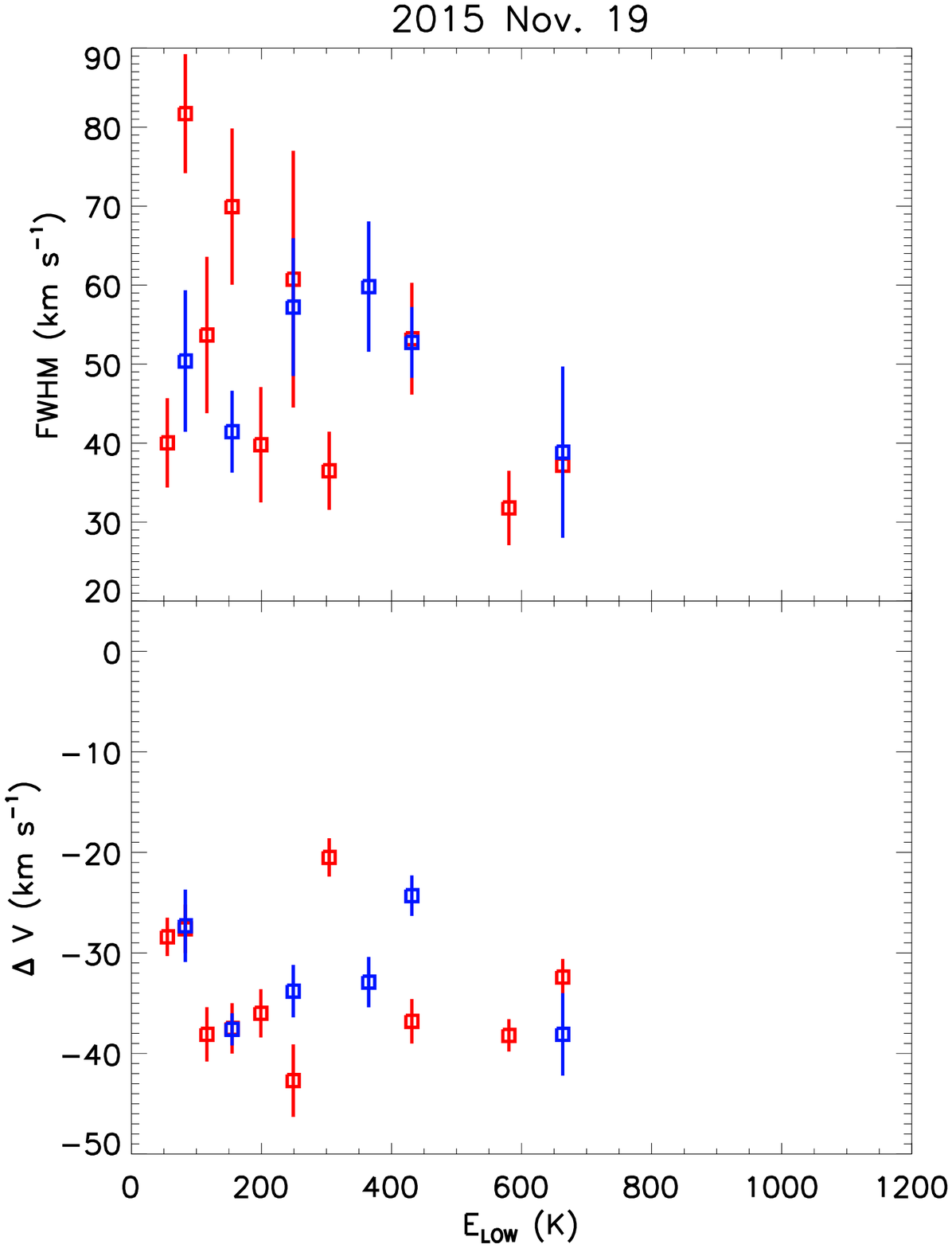}
\caption{Line width (top) and central velocity relative to the source velocity (bottom) of the medium-width CO rovibrational absorption spectra ($v=0\rightarrow2$) for IRAS~04239+2436. The symbols are the same as those in Figure~\ref{fig:abs-i03-1}. }
\label{fig:abs-i04B-0}
\end{figure*}

Now, we analyze the \textit{medium-width}  component toward IRAS~04239+2436. We do not discuss the weaker medium-width component toward IRAS~03445+3242.
 According to the rotation diagram (Figure~\ref{fig:abs-i04B}), the shallow medium-width component in IRAS~04239+2436 is produced by gas with a temperature of $\sim$400~K and a CO column density of $\sim$2$\times$10$^{19}$~cm$^{-2}$. This gas component has a similar gas temperature to the warm gas associated with the deep \textit{spectrally unresolved} component.
 If we adopt the disk properties derived from the modeling of the CO overtone band emission, this component is located within the inner few AU of the disk.
  However, its CO column density is larger by a factor of 3, and  its velocity is more shifted to the blue than the \textit{spectrally unresolved} component (Figure~\ref{fig:abs-i04B-0}).
In addition, this gas component associated with the shallow medium-width absorption measured in 2015 has its center  $\sim$20~km~s$^{-1}$ to the blue of where it was the previous year (see the right panel of Figure~\ref{fig:co_line_profile1} and the lower right panel of Figure~\ref{fig:abs-i04B-0}), while the gas temperature and column density are nearly constant.

\section{DISCUSSION}\label{sec:discussion}
\begin{figure*}
\includegraphics[width=0.7 \textwidth]{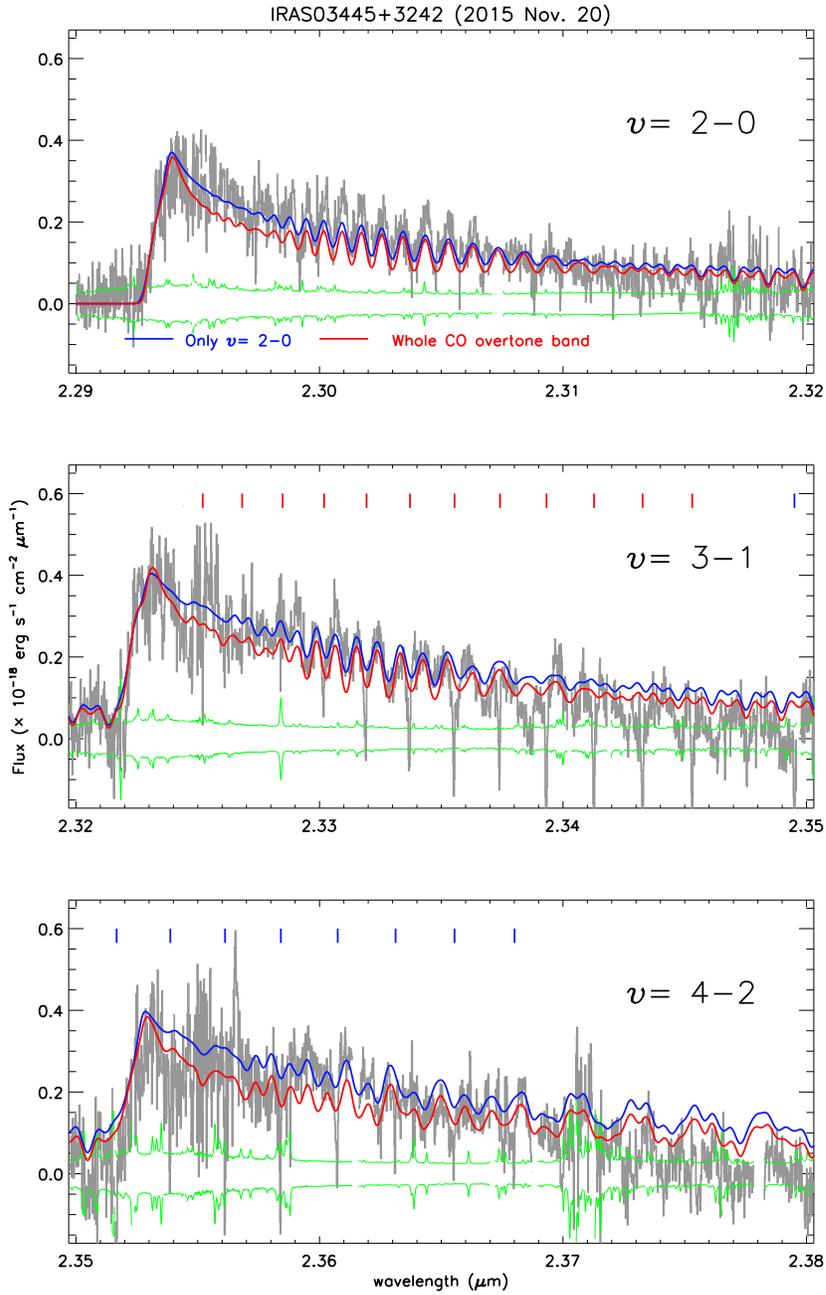}
\caption{Same as Figure~\ref{fig:co_om-i03-1} except for the CO overtone band emission spectra of IRAS~03445+3242 observed on 20 Nov. 2015.}
\label{fig:co_om-i03-2}
\end{figure*}

 In our analysis, we assumed that CO gas in LTE emits the CO overtone band. However, at low density \citep{Najita1996} or if there is significant UV/IR fluorescence \citep{Thi2013}, the CO excitation  can violate the LTE condition. Support for the relevance of non-LTE effects comes from the observation that the rotational and vibrational temperatures of CO fundamental lines are different in HAeBe star disks \citep{van der Plas2015}. To test for non-LTE effects, we tried to find the best-fit model (blue lines in Figures~\ref{fig:co_om-i03-1}--\ref{fig:co_om-i04-2}) only for the first overtone band emission ($v$=~2--0; 2.29-2.32~$\mu$m) and compared it with the best-fit model (red lines in the same figures) derived from the whole spectral range covered by our observations (2.29--2.40 $\mu$m), which includes also the 3--1,  4--2, and 5--3 bands.
 For IRAS~03445-3242, the best-fit models only for $v=$~2--0 have a higher $T_{\rm in}$ and have a better fit in the long-wavelength region of the $v$=~2--0 bandhead (2.294-2.30~$\mu$m) compared to the models for all four CO overtone bands (see top panel of Figures~\ref{fig:co_om-i03-1} and \ref{fig:co_om-i03-2}).
This discrepancy of the two best-fit models in $T_{\rm in}$ implies a (vibrationally) subthermal excitation, and the disk could have relatively low densities in the CO emitting region. Thus, non-LTE effects might be important in IRAS~03445+3242. On the other hand, for IRAS~04239-2436, a similar gas temperature can fit both the $v$=2--0 overtone part only and the whole four overtones, indicating a thermalized disk.

 The CO overtone band spectra of our two Class I sources show that the disks of the Class I sources are similar to more evolved disks. 
 The observations and models of CTTSs and HAeBe stars show that the CO overtone band emission originates in a gaseous Keplerian disk within 0.15~AU \citep{Berthoud2007,Eisner2014} and that  CO overtone emission is coincident with the dust continuum emission. 
 According to our modeling, the CO overtone emission in our sources is radiated from a gaseous Keplerian disk with a gas temperature ranging from $\sim$3500 to $\sim$2500~K and a projected rotational velocity of 60--75 km~s$^{-1}$ at the innermost edge.  Considering photometric magnitudes, inclinations, and the visual extinctions of our targets, this result also suggests that the CO gas arises closer to the star than the inner edge of the dust disks. 
The CO molecule could survive from the photodissociation even in the dust-free inner gas disk. As argued by \citet{Thi2005}, the CO column density of the best-fit model for CO bandhead emission, which originates from dust-free inner hot disk, is  higher than the value required to be self-shielded \citep[$N$(CO)$\sim10^{15}$~cm$^{-2}$,][]{van Dishoeck1988}. In addition,
a high CO formation rate (C + OH $\rightarrow$ CO + H) at the high gas density and temperature could replenish the CO molecules.

We modeled the  \textit{spectrally unresolved} CO vibrational absorption ($v= 0\rightarrow 2$) features with warm and cold components in Section~\ref{ss:out}. In a realistic case, the gas temperature along the line of sight might vary continuously. With that in mind, we note that the positive curvatures in the excitation diagrams can also be reproduced by a power-law temperature distribution \citep{Neufeld2012,Lee2014}.  We therefore also fit the rotation diagram with a power-law temperature distribution ($dN \propto T^{-\alpha} dT$) with the free parameters of the lower and upper temperature limits and the power index $\alpha$ in LTE (see bottom panels in Figures~\ref{fig:abs-i03-0} and \ref{fig:abs-i04-0}). The best-fit models have $\chi^2$ values similar to those in the two-component models. Both models have similar column densities. The lower limit of the gas temperature tends to have a lower gas temperature than 10 K. The upper limit of gas temperature is above 1000 K, indicating that the absorbing layer is connected to the inner disk where the CO overtone band emission originates. 

Narrow CO overtone absorption lines are also observed toward other Class~I sources and CTTSs \citep{Horne2012,Smith2009,Smith2015}. 
Independent of the inclination, the CTTSs show only a warm component which has a gas temperature higher than 250~K and a CO column density of 10$^{18}-10^{19}$~cm$^{-2}$ and  is located in flared disks \citep{Horne2012}. Exceptionally, the absorption toward VV~CrA shows a cold component ($T_{\rm rot}$= 16~K and N(CO)=~1.4$\times10^{18}$~cm$^{-2}$) as well as the warm component, and \citet{Smith2009} conclude that the cold component arises from the outer disk near the midplane. The warm ($\sim$80-250~K) and cold ($\sim$5-30~K) components with the CO column densities of $\sim$10$^{18}$~cm$^{-2}$ are located at the inner and the outer parts of the envelope, respectively, in Class I sources \citep{Smith2015}.
The warm component in the CTTSs has a higher gas temperature than that in the Class I sources, while both have similar CO column densities. Our sources show similar CO column densities to previous work. If we adopt a gas temperature of 250~K as a boundary of  the flared disk and the inner envelope, the warm component of IRAS~03445+3242 and IRAS~04239+2436 arises in the inner envelope and the flared disk, respectively. The cold component in both sources could be located in the envelope and/or in the outer disk; the line of sight toward the protostar and the hot inner disk should pass through both the envelope and outer disk because these sources have high inclinations.

The \textit{medium-width} absorption components are clearly detected in IRAS~04239+2436 with a similar gas temperature to that of the warm component associated with the spectrally unresolved absorption lines, implying that both components could be located at similar radii (a few AU) of a flared disk. 
 However, the difference in line-width between the spectrally unresolved ($\sim$8~km~s$^{-1}$) and medium-width ($>$ 30~km~s$^{-1}$) components needs other interpretations; the projected Keplerian rotation velocity at a few AU can produce the line width of $\sim$20~km~s$^{-1}$ at the maximum when the continuum source size is similar to a few AU.   
In addition, an axisymmetric Keplerian disk cannot explain why the velocity  of the medium-width absorption is blueshifted compared to that of the spectrally unresolved absorption. To explain the blueshifted medium-width absorption, an asymmetric feature of the protoplanetary disk is required (such as a warped disk).

The \textit{medium-width} absorption components could also originate from an outflow. The CO fundamental emission toward some Class I sources has a blueshifted (up to $\sim$10~km~s$^{-1}$) medium line width (10--50~km~s$^{-1}$) component, indicating a slow outflow \citep{Herczeg2011}. This CO fundamental emission shows a gas temperature (300--400~K) and a CO column density ($\sim$10$^{19}$~cm$^{-2}$) similar to those of our \textit{medium-width} absorption component. However, the high blueshifted  velocity (-35~km~s$^{-1}$) of the last epoch observation requires an additional explanation.

\section{SUMMARY}
We analyzed the high resolution near-IR spectra observed with IGRINS to study disks in the early phase of low-mass star formation.
The hot inner disks of two embedded Class I objects have been detected in CO overtone emission as well as in features of Na~I and Ca~I. A simple thin Keplerian disk model with the temperature of 3500 K and a rotation velocity ($V\sin i$) of $\sim$60--85 km~s$^{-1}$ at the innermost radius fits the observations well. 
 We also see CO rovibrational absorption spectra of the first overtone band  where the distribution of $J$-state populations is consistent either with two temperature components ($\sim$150--480~K and  20--30~K) model or with a power-law temperature model.  The warm gas exists at a radius of  a few AU, if we adopt the same disk as used to fit the CO overtone band emission, implying that the CO rovibrational absorption spectra can trace the warm gas in a few AU of the disk or innermost part of the envelope. Na~I and Ca~I emission lines are originated from the same Keplerian disk producing the CO overtone band emission although the atomic lines show asymmetric line features because of the photospheric absorption lines. Multi-epoch observations toward IRAS~03445+3242 show that the red peaks of CO overtone bandhead, Na I , and Ca I emission lines decrease in one year implying non-axisymmetric features of protoplanetary disk.
On the other hand,  for IRAS~04239+2436, only atomic lines show the rapid decline of the flux of  blueshifted velocity component, which might be due to a dark spot on the protostar. In conclusion, IGRINS is a powerful instrument to study the sub-AU scale hot and AU-scale warm disks simultaneously.

\acknowledgments
 This work was supported by the Basic Science Research Program (grant No. NRF-2015R1A2A2A01004769) and the BK21 plus program through the National Research Foundation (NRF) funded by the Ministry of Education of Korea. 
 This research was also supported by the Korea Astronomy and Space Science Institute under the R\&D program (Project No. 2015-1-320-18) supervised by the Ministry of Science, ICT and Future Planning.
This work used the Immersion Grating Infrared Spectrograph (IGRINS) that was developed under a collaboration between the University of Texas at Austin and the Korea Astronomy and Space Science Institute (KASI) with the financial support of the US National Science Foundation under grant AST-1229522, of the University of Texas at Austin, and of the Korean GMT Project of KASI.

\bibliographystyle{aa}
\bibliography{biblio}

\begin{deluxetable}{cccc}
\tablecaption{Observation log \label{tb:obslog}}
\tablehead{\colhead{Source} &
	        \colhead{Date} &
	        \colhead{Total integrated time (s)} &
	        \colhead{Calibrator} }
\startdata
IRAS~03445+3242 & 2015 Jan. 21 & 2400 & 18 Ori  \\
                & 2015 Nov. 20 & 5400 & k tau \\
IRAS~04239+2436 & 2014 Nov. 27 & 2400 & 18 Ori \\
	 			  & 2014 Nov. 28 & 2400 & 18 Ori \\
	 			  & 2015 Nov. 19 & 3600 & k tau       
\enddata
\end{deluxetable}

\begin{deluxetable}{ccccccccc}
\tablewidth{0pt}
\rotate
\tablecaption{Source list and parameters \label{tb:target} }
\tablehead{  &\colhead{R.A. (J2000)\tablenotemark{1}}
			 &\colhead{Dec. (J2000) \tablenotemark{1}}
			 &\colhead{$K_{\rm mag}$\tablenotemark{1} }
			 &\colhead{$V_{\rm LSR}$\tablenotemark{2, 3}}
			 &\colhead{distance\tablenotemark{4, 5} }
			 &\colhead{inclination\tablenotemark{5,6}}
			 &\colhead{$A_{\rm V}$\tablenotemark{a}}
			 &\colhead{$L_{\rm bol}$\tablenotemark{1} }\\
			 
			 &\colhead{}
			 &\colhead{}
			 &\colhead{(mag)}
			 &\colhead{(${\rm km~s^{-1}}$)}
			 &\colhead{(pc)}
			 &\colhead{($^\circ$)}
			 &\colhead{(mag)}
			 &\colhead{($L_\odot$)}
			 }
\startdata
IRAS~03445+3242 &03:47:41.60 &32:51:43.8  &11.49 &10.27 $\pm$ 0.06  & 240 
                &77 $\pm$ 10 & 22.8 & 3.8 \\
IRAS~04239+2436 &04:26:56.30 & 24:43:35.3  & 10.22 & 6.54 $\pm$ 0.02 & 140
			    & 80 $\pm$ 5 & 19.5  &  1.1 
\enddata
\tablenotetext{a}{Visual extinction is derived from the silicate features at 9.7 $\mu$m (see Section~\ref{sss:co}).}
\tablerefs{(1) \citealt{Connelley2010}; (2) \citealt{Emprechtinger2009}; (3) \citealt{Fuller2002}; (4) \citealt{Zapata2014}; (5) \citealt{Davis2003} ; (6) \citealt{Arce2001}.}
\end{deluxetable}

\begin{deluxetable}{ccccccc}
\tablewidth{0pt}
\rotate
\tablecaption{The best-fit model parameters for the whole four CO overtone bands emission \label{tb:bestfit} }
\tablehead{    & \multicolumn{2}{c}{IRAS~03445+3242} & &
            \multicolumn{3}{c}{IRAS~04239+2345} \\
            \cline{2-3} \cline{5-7} \\
     & 2015 Jan. 21 & 2015 Nov. 20 & &2014 Nov. 27 & 2014 Nov. 28 & 2015 Nov. 19 }
\startdata
\multicolumn{3}{l}{The best-fit free parameters} \\
\hline
      $V_{\rm in}$             (km s$^{-1}$)  & 60  & 70 && 75 &85 & 75 \\
      $\Delta V_{\rm FWHM}$    (km s$^{-1}$)  & 45  & 50 && 50&45 & 50 \\
      $T_{\rm in}$              (K)           & 3000 &  3500 &&3500 &3500 & 3750\\
      $T_{\rm out}$             (K)          & 2500 & 3000 &&2500 &2500 & 3000 \\
      $N_{\rm in}$  ($\times$ 10$^{20}$ cm$^{-2}$) & 4.6 & 10.0 && 10.0 & 10.0 & 4.6  \\
\hline
\multicolumn{3}{l}{The derived parameters}\\
\hline
      $r_{\rm in}$            ($\times$ 10$^{-2}$ AU) & 3.9$^{+4.2}_{-1.0}$  
      												 & 2.4$^{+2.5}_{-0.5}$&                
                                                     & 2.2$^{+0.8}_{-0.4}$
                                                     & 2.1$^{+0.9}_{-0.4}$
                                                     & 3.1$^{+1.3}_{-0.6}$\\
      $r_{\rm out}$            ($\times$ 10$^{-2}$AU) & 5.7$^{+6.1}_{-1.4}$
                                                     & 3.2$^{+3.4}_{-074}$ &
                                                     & 4.2$^{+1.7}_{-0.8}$
                                                     & 4.1$^{+1.7}_{-0.7}$
                                                     & 4.8$^{+2.0}_{-0.9}$\\
      $M_{\rm star}$           ($M_\odot$)            & 0.17$^{+0.16}_{-0.03}$
                                                     & 0.14$^{+0.13}_{-0.02}$&
                                                     & 0.14$^{+0.05}_{-0.02}$
                                                     & 0.18$^{+0.07}_{-0.03}$
                                                     & 0.20$^{+0.08}_{-0.03}$
\enddata
\tablecomments{The uncertainties of the free parameters are estimated by the bootstrap method, which are lower than or comparable to the step size (see Section~\ref{sss:co}). The derived parameters are obtained from the best-fit free parameters for a given distance, inclination, and visual extinction in Table~\ref{tb:target}. The uncertainties of the derived parameters mainly come from the uncertainty of the inclination.}
\end{deluxetable}

\begin{deluxetable}{ccccccc}
\tablewidth{0pt}
\rotate
\tablecaption{The best-fit model parameters only for the CO $\upsilon=$ 2--0 band emission \label{tb:bestfit1} }
\tablehead{    & \multicolumn{2}{c}{IRAS~03445+3242} & &
            \multicolumn{3}{c}{IRAS~04239+2345} \\
            \cline{2-3} \cline{5-7} \\
     & 2015 Jan. 21 & 2015 Nov. 20 & &2014 Nov. 27 & 2014 Nov. 28 & 2015 Nov. 19 }
\startdata
\multicolumn{3}{l}{The best-fit free parameters} \\
\hline
      $V_{\rm in}$             (km s$^{-1}$)  & 65  & 90 && 80 &80 & 75 \\
      $\Delta V_{\rm FWHM}$    (km s$^{-1}$)  & 45  & 50 && 50&45 & 50 \\
      $T_{\rm in}$              (K)           & 4000 &  5750 &&3000 &3250 & 3500\\
      $T_{\rm out}$             (K)           & 3000 & 3000 &&2000 &2500 & 3000 \\
      $N_{\rm in}$  ($\times$ 10$^{20}$ cm$^{-2}$) &21.5 & 46.4 && 1.0 & 10.0 & 2.2  \\
\hline
\multicolumn{3}{l}{The derived parameters}\\
\hline
      $r_{\rm in}$            ($\times$ 10$^{-2}$ AU) & 1.5$^{+1.6}_{-0.4}$  
      												 & 0.5$^{+0.5}_{-0.1}$&                
                                                     & 6.7$^{+2.7}_{-1.2}$
                                                     & 2.5$^{+1.0}_{-0.5}$
                                                     & 5.5$^{+2.3}_{-1.0}$\\
      $r_{\rm out}$            ($\times$ 10$^{-2}$AU) & 2.6$^{+2.8}_{-0.6}$
                                                     & 1.8$^{+1.9}_{-0.4}$ &
                                                     & 15.0$^{+6.2}_{-2.7}$
                                                     & 4.3$^{+1.8}_{-0.8}$
                                                     & 7.6$^{+3.1}_{-1.4}$\\
      $M_{\rm star}$           ($M_\odot$)            & 0.07$^{+0.07}_{-0.01}$
                                                     & 0.05$^{+0.05}_{-0.01}$&
                                                     & 0.50$^{+0.19}_{-0.07}$
                                                     & 0.19$^{+0.07}_{-0.03}$
                                                     & 0.36$^{+0.14}_{-0.05}$
\enddata
\tablecomments{Same as Table~\ref{tb:bestfit} but only for CO $v$=2--0 band.}
\end{deluxetable}


\clearpage

\clearpage

\clearpage

\end{document}